\documentclass[aps,prd,superscriptaddress,floatfix,nofootinbib,preprintnumbers,eqsecnum,twocolumn]{revtex4-2}

\pdfoutput=1
\usepackage{amssymb,amsfonts,amsmath,graphicx}
\usepackage{xcolor}
\usepackage{orcidlink}
\usepackage[english]{babel}
\usepackage[T1]{fontenc}
\usepackage{amsmath}
\usepackage{slashed}
\usepackage{booktabs}
\usepackage{listings}
\usepackage[utf8]{inputenc}
\usepackage{circuitikz}
 \hypersetup{ 
   colorlinks,
   linkcolor={blue!80!black},
   citecolor={blue!70!black},
   urlcolor={blue!70!black}
 }
\usepackage{mathtools}
\usepackage{footmisc}

\usepackage{placeins}

\usepackage{amssymb,amsmath,epsfig,longtable}

\numberwithin{equation}{section}

\newcommand{\bean}{\begin{eqnarray*}}
\newcommand{\eean}{\end{eqnarray*}}

\newcommand{\fref}[1]{Figure~\ref{#1}}

\newcommand{\rk}{\mathop{{\rm rk}}}

\newcommand{\cO}{{\cal O}}
\newcommand{\cN}{{\cal N}}

\newcommand{\cA}{{\cal A}}
\newcommand{\cB}{{\cal B}}
\newcommand{\cC}{{\cal C}}

\newcommand{\cV}{{\cal V}}

\def\cjn1{{\cA, \cC^*\otimes \wedge^j \cN^*}}
\def\bjn1{{\cA, \cB^*\otimes \wedge^j \cN^*}}
\def\vjn1{{\cA, \cV^*\otimes \wedge^j \cN^*}}
\def\cjn2{{\cA, \cC\otimes \wedge^j \cN^*}}
\def\bjn2{{\cA, \cB\otimes \wedge^j \cN^*}}
\def\vjn2{{\cA, \cV\otimes \wedge^j \cN^*}}

\newcommand{\be}{\begin{equation}}
\newcommand{\ee}{\end{equation}}
\newcommand*{\nnbe}{\begin{equation}}
\newcommand*{\nnee}{\end{equation}}
\newcommand{\bea}{\begin{eqnarray}}
\newcommand{\eea}{\end{eqnarray}}
\newcommand{\ba}{\begin{align}}
\newcommand{\ea}{\end{align}}
\newcommand{\bi}{\begin{itemize}}
\newcommand{\ei}{\end{itemize}}

\newsavebox{\overlongequation}

\DeclarePairedDelimiter\set\{\}

\usepackage{hyperref}

\begin{document}
\title{Enumerating Calabi-Yau Manifolds:\\Placing bounds on the number of diffeomorphism classes in the Kreuzer-Skarke list}



\author{Aditi Chandra \orcidlink{0000-0002-0798-6904}}
\email[]{aditi.chandra@balliol.ox.ac.uk}
\affiliation{Balliol College, University of Oxford, Broad St, Oxford OX1 3BJ, UK}

\author{Andrei Constantin \orcidlink{0000-0002-0861-5363}}
\email[]{andrei.constantin@physics.ox.ac.uk}
\affiliation{Rudolf Peierls Centre for Theoretical Physics, University of Oxford, Parks Road, Oxford OX1 3PU, UK}

\author{Kit~Fraser$\text{-}\!\!\!$~Taliente~\orcidlink{0009-0008-8037-6855}}
\email[]{cristofero.fraser-taliente@physics.ox.ac.uk}
\affiliation{Rudolf Peierls Centre for Theoretical Physics, University of Oxford, Parks Road, Oxford OX1 3PU, UK}

\author{Thomas R.~Harvey \orcidlink{0000-0002-4990-4778}}
\email[]{thomas.harvey@physics.ox.ac.uk}
\affiliation{Rudolf Peierls Centre for Theoretical Physics, University of Oxford, Parks Road, Oxford OX1 3PU, UK}

\author{Andre Lukas \orcidlink{0000-0003-4969-0447}}
\email[]{andre.lukas@physics.ox.ac.uk}
\affiliation{Rudolf Peierls Centre for Theoretical Physics, University of Oxford, Parks Road, Oxford OX1 3PU, UK}


\begin{abstract}
The diffeomorphism class of simply connected smooth Calabi-Yau threefolds with torsion-free cohomology is determined via certain basic topological invariants: the Hodge numbers, the triple intersection form, and the second Chern class. In the present paper, we shed some light on this classification by placing bounds on the number of diffeomorphism classes present in the set of smooth Calabi-Yau threefolds constructed from the Kreuzer-Skarke list of reflexive polytopes up to Picard number six. The main difficulty arises from the comparison of triple intersection numbers and divisor integrals of the second Chern class up to basis transformations. By using certain basis-independent invariants, some of which appear here for the first time, we are able to place lower bounds on the number of classes. Upper bounds are obtained by explicitly identifying basis transformations, using constraints related to the index of line bundles. Extrapolating our results, we conjecture that the favourable entries of the Kreuzer-Skarke list of reflexive polytopes leads to some $10^{400}$ diffeomorphically distinct Calabi-Yau threefolds.

\end{abstract}
\maketitle		


\section{Introduction and Summary}

The purpose of this paper is to distinguish the diffeomorphism class of smooth, simply connected Calabi-Yau threefolds, defined as compact K\"ahler threefolds $X$ with trivial canonical bundle and vanishing $H^1(X,\cO_X)$. We assume that certain basic topological invariants are known, namely the Hodge numbers, the symmetric trilinear intersection form, and the second Chern class given explicitly relative to an integral basis of $H^2(X,\mathbb Z)$. The key problem addressed here is to decide when two sets of intersections forms and second Chern classes are the same up to a basis transformation. 

The pair of Hodge numbers $(h^{1,1}(X), h^{2,1}(X))$ specifies what is known in the literature as the {\itshape topological type} of a Calabi-Yau threefold $X$. 
Certainly, if two Calabi-Yau threefolds are homeomorphic to each other, their topological types must agree. The converse, however, is not true. Two Calabi-Yau threefolds of the same topological type may differ with respect to other invariants, such as the trilinear form and second Chern class. For simply connected manifolds with torsion-free cohomology, Wall's theorem~\cite{WALL1966} states that the isomorphism class of the system of invariants mentioned above, including the Hodge numbers, specifies uniquely the {\itshape diffeomorphism type}. Two non-singular diffeomorphic Calabi-Yau threefolds are also deformation equivalent~\cite{Tian:1987, TodorovL1989}, that is, their {\itshape homotopy types} agree, provided that the diffeomorphism types agree;  however, this fails to hold in general~\cite{Gross:1997}. 

The question addressed in this paper is of importance to both pure mathematics and physics. Since CY manifolds form one of the building blocks in the classification of algebraic varieties up to birational isomorphisms, their classification is an important and open problem in algebraic geometry. From a differential geometric perspective, the question of classifying CY manifolds up to diffeomorphisms is natural. In dimensions one, all CY manifolds, that is, genus-one curves, are diffeomorphic to each other. The same is true in dimension two: all CY2 surfaces, a.k.a. K3 surfaces, are diffeomorphic as smooth 4-manifolds. In dimension three, the picture is much more diverse and, to a large extent, unresolved. For instance, it is not known whether the number of distinct topological types is finite, except in the case of elliptically-fibered CY threefolds, for which the answer is positive~\cite{Grassi1991, Gross1994}. It is also not known what kind of trilinear intersection forms can occur on CY threefolds, unlike the case of K3 surfaces, where the possible intersection forms are classified by the even self-dual lattices of signature $(3,19)$. Another interesting question is whether the number of distinct Hodge pairs that can arise for a given isomorphism class of triple intersection numbers and divisor integrals of the second Chern class is bounded or not. It turns out that this is true for CY threefolds containing no rigid non-movable surfaces~\cite{Wilson2017BoundednessCYs}, however, it is unclear whether this statement generalises.

In physics, the classification of CY threefolds is closely related to the classification of supergravity theories derived from string theory or M-theory. For instance, the low-energy limit of M-theory compactified on a CY threefold $X$ is a five-dimensional $\mathcal N=1$ supergravity theory with $h^{1,1}(X)-1$ vector multiplets and $h^{2,1}(X)+1$ hypermultiplets. The triple intersection numbers of $X$ determine the field-space metric for the vector multiplets and certain Chern-Simons couplings, while the integrated second Chern class fixes certain higher-curvature terms. A recent research avenue in this context, known as the swampland program~\cite{Vafa:2005ui, Palti:2019pca, vanBeest:2021lhn}, aims to study certain classes of field theories derived from string and M-theory in order to extrapolate their common features to general statements about field theories that can be embedded into quantum gravity in the ultraviolet. The triple intersection numbers also arise in the computation of A-model Yukawa couplings, given by an infinite series that collect contributions from holomorphic maps of all possible degrees from genus-0 curves into $X$. Such infinite series correspond to the `quantum intersection numbers' of the manifold, which reduce to their classical counterparts in the limit when $X$ has large radius. 

Many constructions of CY threefolds have been developed both in algebraic and complex geometry. The largest set of known examples, arises from Batyrev's construction in the context of mirror symmetry~\cite{Batyrev:1993dm}. Given a four-dimensional reflexive polytope $\Delta$, the construction associates with $\Delta$ a Gorenstein toric Fano variety and a generic anticanonical section therein, whose crepant resolution produces a CY threefold $X$. The Hodge numbers of $X$ are determined by the combinatorial data of the polytope, while the triple intersection numbers and the divisor integrals of the second Chern class are computed from the triangulation of $\Delta$ associated with the crepant resolution. The classification of four-dimensional reflexive polytopes undertaken by Kreuzer and Skarke~\cite{kreuzer2000complete} includes $473,800,776$ polytope isomorphism classes, with $30,108$ distinct pairs of Hodge numbers. The large degeneracy of the Hodge numbers can be partly understood as a consequence of the K3 fibration structures that abound among the Kreuzer-Skarke (KS) manifolds. A distinctive feature of these fibrations in that the K3 polyhedron is contained in the four-dimensional polytope as a slice, dividing it into two parts, a top and a bottom. The Hodge numbers satisfy an additivity relation with respect to the operation of assembling reflexive polytopes from tops~\cite{Candelas:2012uu}, which is largely the source of the Hodge number degeneracy. Another feature of the manifolds constructed from the KS list is that these populate a region with $h^{1,1}(X)+h^{1,2}(X)\geq 22$. The complementary region $h^{1,1}(X)+h^{1,2}(X)< 22$ is much less populated, and all known examples arise from other constructions~\cite{Candelas:2016fdy}. The KS list includes 16 polytopes that lead to non-simply connected manifolds~\cite{Batyrev:2005jc}. Throughout this work, these have been removed from the KS list as in this case Wall's theorem does not apply.

The number of diffeomorphism types present in the set of CY threefolds derivable from the KS list of four-dimensional reflexive polytopes is not known. In principle, this can be very large, since the number of triangulations grows exponentially with $h^{1,1}(X)$ and the KS list includes examples for which $h^{1,1}(X)$ is as large as $491$. An upper bound on the number of fine, regular, star triangulations of four-dimensional polytopes was given in Ref.~\cite{Demirtas:2020dbm} as $1.53\times 10^{928}$. In the same paper, a weak upper bound on the number of diffeomorphism classes of CY threefolds derivable from the KS list was given as $1.65 \times 10^{428}$. Both of these bounds are dominated by the single largest polytope with the largest number of points, which is associated to manifolds $X$ with $h^{1,1}(X)=491$. The number of distinct Hodge number pairs, namely 30,108, gives a weak lower bound. Our aim here will be to provide bounds for the number of diffeomorphism types present in this class of CY threefolds up to Picard number 6. The main difficulty arises from the comparison of triple intersection numbers and divisor integrals of the second Chern class up to basis transformations on the second integral cohomology $H^2(X,\mathbb Z)$. 

To be more specific, we denote the triple intersection form $\kappa$ on $H^2(X,\mathbb Z)$ by $\kappa(A,B,C) = A\cdot B \cdot C$, where $A,B,C\in H^2(X,\mathbb{Z})$, and the second Chern class of $X$ by $c_2(X)$. Let us consider two smooth, simply connected CY threefolds $X$ and~$X'$ with torsion-free cohomology, Hodge numbers $h:=h^{1,1}(X)$, $h':=h^{1,1}(X')$ and intersection form and second Chern classes denoted by $\kappa$, $c_2(X)$ and $\kappa'$, $c_2(X')$, respectively.
Further, we introduce integral bases\footnote{For CY hypersurfaces in Gorenstein toric Fano varieties, if the polytope is \emph{favourable} in the sense of Ref.~\cite{Demirtas:2018akl}, an integral basis on the second cohomology of the threefold can be obtained by restriction from the ambient variety.}
$(D_i)$, where $i=1,\ldots ,h$, of $H^2(X,\mathbb{Z})$, and $(D_i')$, where $i=1,\ldots ,h'$, of $H^2(X',\mathbb Z)$. Relative to these bases, we can define the intersection numbers and components of the second Chern class by
\begin{equation}
    \begin{aligned}
   d_{ijk} &= \kappa(D_i,D_j, D_k), & \quad d'_{ijk} &= \kappa'(D'_i,D'_j,D'_k),\\
    c_{2,i} &= c_2(X) \cdot D_i, & \quad c'_{2,i} &= c_2(X') \cdot D_i'.
    \end{aligned}
\end{equation}

Let us assume that $H^2(X,\mathbb Z)\simeq H^2(X',\mathbb Z)$ (so that, in particular, $h=h'$) and $H^3(X,\mathbb Z)\simeq H^3(X',\mathbb Z)$. Then, the two manifolds are diffeomorphic if and only if there exists an invertible integral transformation 
\begin{equation}
P:H^2(X,\mathbb Z)\rightarrow H^2(X',\mathbb Z)~,
\end{equation}
such that, relative to the given bases,
\begin{equation}\label{eqns_for_P}
\begin{aligned}
c'_{2,i} &~=~ {P_i}^r c_{2,r}~, \\
d_{ijk}'&~=~ {P_i}^r{P_j}^s{P_k}^t d_{rst}~,
\end{aligned}
\end{equation}
where $P\in \operatorname{GL}(h,\mathbb{Z})$ is an invertible $h\times h$ matrix with integer entries.
A somewhat simpler question is to determine whether a solution for $P$ exists over the real numbers. Of course, if the answer turns out to be negative, then the same must be true over the integers and the two manifolds cannot be diffeomorphic to each other. A classification of the triple intersection forms in Picard number two up to real basis transformations has been carried out in Ref.~\cite{BrodieGeodesicExtKahler}, revealing four different classes. Attempting a similar classification in higher Picard number is non-trivial. However, the problem of solving \eqref{eqns_for_P} for a real matrix $P$ can be turned into an optimisation problem, as discussed in Ref.~\cite{CandelasHowManyCICYs}.

The computational complexity of an exhaustive search for integral transformations satisfying \eqref{eqns_for_P} with the entries of $P$ in a range $[-k_{\rm max},k_{\rm max}]$ is $\mathcal O(k_{\rm max}^{h^2})$. In Section~\ref{sec:search_alg}, we introduce an alternative search algorithm with computational complexity ${\mathcal O}(k_{\rm max}^h)$
and, in Section~\ref{sec:Results}, we apply this algorithm to obtain an upper bound on the number of diffeomorphism classes present in the list of KS CY threefolds up to Picard number six. For each $h$, the value of $k_{\rm max}$ is increased until either the bound stabilises or the computational time becomes impractical (for example, for $h=6$ we did not attempt to extend the search beyond $k_{\rm max}=5$.). Even in the cases where the upper bound did not stabilise, the dependence on $k_{\rm max}$ strongly indicates that the actual number of diffeomorphism classes is very close to the obtained bound. 

The existence of integral transformation satisfying \eqref{eqns_for_P} implies diffeomorphism equivalence. On the other hand, necessary criteria for equivalence can be derived from certain basis-independent quantities, including the Hodge numbers, the GCD-invariants found by H\"ubsch in Ref.~\cite{hubschCYB}, as well as other invariants, to be discussed in Section~\ref{sec:LowerBound}. The latter invariants are derived from polynomial invariant theory and, to our knowledge, are discussed here for the first time, at least in the context of CY threefolds. Applied to the list of KS CY threefolds up to Picard number six, the invariants define a number of distinct classes, which leads to a lower bound on the total number of diffeomorphism classes. Moreover, these invariants inform the application of the proposed search algorithm, since the existence of transformations satisfying \eqref{eqns_for_P} only has to be tested for pairs of manifolds for which all invariants coincide. Clearly, the fewer elements we have in the classes of manifolds with the same invariants, the more efficient the search algorithm is. 

A different set of invariants arises from the study of limiting mixed Hodge structures (approximately speaking, these describe the behaviour of the Hodge decomposition of the third cohomology at the boundaries of the complex structure moduli space), as recently pointed out in Ref.~\cite{Grimm:2019bey}. However, we will not make use of these latter invariants in our analysis. 
Attempts to machine learn certain basis-independent invariants have appeared in Ref.~\cite{TaylorJejjala}.

The bounds on the number of diffeomorphism classes present in the list of KS CY threefolds up to Picard number six are discussed in Section~\ref{sec:Results}, and summarised in Table~\ref{tab:Results} and Fig.~\ref{fig:TriangEquivPlot}. The lower and upper bounds are close to each other, with a maximal difference of about $5\%$ for Picard number $h=6$. Moreover, plotted on a logarithmic scale as functions of the Picard number $h$, the bounds align on almost straight lines. Consequently, the number of diffeomorphism classes contained in the favourable entries of the KS CY threefolds list, at least for $h\leq 6$, follows the simple formula,
\begin{equation}
\label{eq:ndiff}
    n_{\rm diff} = (0.68\pm0.02) e^{(1.87\pm 0.01)h}~,
\end{equation}
which amounts to an approximate scaling factor of $6.5$ per unit increment in $h$. By comparison, the number of fine, regular, star triangulations (FRSTs) for $1\leq h\leq 6$ follows a more precise exponential law:
\begin{equation}
\label{eq:ntr}
    n_{\rm triang} = (0.48698\pm 0.00005) e^{(2.33295\pm 0.00002)h}~,
\end{equation}
which amounts to an approximate scaling factor of $10.3$ per unit increment in $h$.

It is tempting to extrapolate these formulae beyond $h=6$. By summing up the contributions from all Picard numbers in the range $1\leq h \leq 491$, a na\"ive extrapolation leads to an estimate of $(1.6\pm0.1)\times10^{497}$ FRSTs and a total number of diffeomorphism classes between $10^{396}$ and $10^{401}$ in the (favourable) KS-list, both of which fall within the bounds of Ref.~\cite{Demirtas:2020dbm}. These estimates are dominated by the contributions of the single polytope which leads to manifolds $X$ with $h^{1,1}(X)=491$.

\section{A plethora of invariants}\label{sec:LowerBound}
In this section, we introduce a number of diffeomorphism invariants constructed from the second Chern class and the intersection numbers. These invariants are used in Section~\ref{sec:Results} to place lower bounds on the number of CY threefolds which can be constructed from the KS list. In Section~\ref{sec:GCD}, we review the GCD invariants found in Ref.~\cite{hubschCYB}, and, in Section~\ref{sec:TensorPowersAndInvariants}. we generalise these using representation theory of $\operatorname{GL}(h,\mathbb Z)$ in Section~\ref{sec:TensorPowersAndInvariants}. In addition, there are also polynomial invariants, which we introduce and discuss in Section~\ref{sec:PolyInv}. With the exception of the simplest GCD invariants, to the best of our knowledge, none of the others presented here have been used in the context of CY threefolds. Subsection~\ref{sec:InvDiscrimPower} explores the discriminative potential of each of the invariants.

Ideally, one would like to produce a number of invariants that are sufficiently low in computational complexity to be of practical use and at the same time sufficiently powerful to distinguish between diffeomorphically non-equivalent CY threefolds. Below, we construct many invariants that almost completely distinguish between the diffeomorphism classes present in the set of KS CY threefolds up to Picard number $6$. The discussion is regrettably labyrinthine and in order to streamline the presentation we have delegated many of the details to footnotes and appendices. 

The classification of the divisor integrals of the second Chern class up to integral basis transformations is trivial, as vectors in the fundamental of $\operatorname{GL}(h,\mathbb{Z})$ are classified by the GCD of their entries. In order to gain some appreciation for the difficulty of the problem, let us focus for a moment on the classification of the triple intersection numbers alone, that is, we consider symmetric trilinear forms up to integral basis transformations. Since a general result for trilinear forms is not known, we take a step back and recall what is known about the classification of symmetric bilinear forms. 
Sylvester's theorem gives the invariants which fully determine equivalence up to $\operatorname{GL}(h,\mathbb{R})$ transformations, namely the rank and the signature. By specifying an additional (polynomial) invariant, the determinant, equivalence up to $\operatorname{SL}(h,\mathbb{R})$ transformations can be determined. A complete classification of quadratic forms (both definite and indefinite) up to rational equivalence exists, and this makes use of $p$-adic invariants. Restricting further to $\operatorname{SL}(h,\mathbb{Z})$ transformations, more invariants are needed, namely the GCDs of the integers specifying the bilinear form in a basis, as well as other more complex invariants, which have a long history in the literature~\cite{ConwaySloane}. 
In dimension two, Gauss' \emph{Disquisitiones arithmeticae} \cite{GaussBook} gives a complete classification of all quadratic forms. In higher dimensions, the 'spinor genus' introduced by Eichler \cite{eichler1952quadratische} gives a complete classification for indefinite forms and a partial classification for definite forms. There exist algorithms that can completely classify all definite quadratic forms in low dimensions, but these become unworkable~\cite{ConwaySloane} beyond dimension $24$.  

Given these complications, we will not attempt a complete classification of trilinear forms, but rather identify a number of simple and powerful invariants. Unlike for bilinear forms, the number of independent entries in a trilinear form is polynomially larger than the number of entries specifying a basis transformation. This suggests a large number of invariants can be expected for trilinear form.

\subsection{GCDs of intersection numbers and divisor integrals of the second Chern class}\label{sec:GCD}

We begin with a discussion of the simplest GCD invariants identified in Ref.~\cite{hubschCYB}. The starting point is the observation that the GCD of the entries of the symmetric array $d_{ijk}$ is preserved under $\operatorname{GL}(h,\mathbb{Z})$ transformations, and likewise, the GCD of the integers $c_{2,r}$. This is a simple consequence of the generalised Bézout's identity, and the fact that the determinant of the transformation is~$\pm1$ (see Appendix \ref{app:basicGCDinvs} for more details). 

In Ref.~\cite{hubschCYB} H\"ubsch identifies a total of eight GCD invariants. The first four of these depend on either $d_{ijk}$ or $c_{2,r}$ and are given by:
\be
\label{eq:Hubsch:St0to3}
\begin{aligned}
\tilde{S}_0 &= \operatorname{GCD}\left(\left\{c_{2,i} \mid i=1, \ldots, h\right\}\right),\\
\tilde{S}_1 &=\operatorname{GCD}\left(\left\{d_{i j k} \mid i, j, k =1, \ldots, h\right\}\right), \\
\tilde{S}_2 &=\operatorname{GCD}\left(\left\{d_{i i j} \mid i, j=1, \ldots, h\right\}\right. \\ 
 & \cup\left.\left\{2 d_{i j k} \mid i, j, k=1, \ldots, h\right\}\right), \\
\tilde{S}_3 &=\operatorname{GCD}\left(\left\{d_{i i i} \mid i=1, \ldots, h\right\}\right. \\
&  \cup\left\{3\left(d_{i i j} \pm d_{i j j}\right) \mid i, j=1, \ldots, h\right\} \\
& \cup\left.\left\{6 d_{i j k} \mid i, j, k=1, \ldots, h\right\}\right).
 \end{aligned}
 \ee

The remaining four invariants involve tensor products of $d_{ijk}$ and $c_{2,r}$. To simplify notation, it is useful to introduce the symmetric quadrilinear form $b_{ijkl}=c_{(i}d_{jkl)}$, in which case the other four invariants are defined as
\be
\label{eq:Hubsch:St4to7}
\begin{aligned}
\tilde{S}_4 & =\operatorname{GCD}\left(\left\{b_{i j k l} \mid i, j, k, l=1, \ldots, h\right\}\right)~, \\
\tilde{S}_5 & =\operatorname{GCD}\left(\left\{b_{i i j k} \mid i, j, k=1, \ldots, h\right\}\right. \\
&\left.\quad\cup\left\{2 b_{i j k l} \mid i, j, k, l=1, \ldots, h\right\}\right)~, \\
\tilde{S}_6 & =\operatorname{GCD}\left(\left\{b_{i i i j} \mid i, j=1, \ldots, h\right\}\right.\\
&\quad\cup\left\{3\left(b_{i i j k} \pm b_{i j j k}\right) \mid i, j, k=1, \ldots, h\right\} \\
& \left.\quad\cup\left\{6 b_{i j k l} \mid i, j, k, l=1, \ldots, h\right\} \right)~,\\
\tilde{S}_7 & =\operatorname{GCD}\left(\left\{b_{i i i i} \mid i=1, \ldots, h\right\}\right. \\
&\quad\cup\left\{6 b_{i i j j} \pm 4\left(b_{i i i j} \pm b_{i j j j}\right) \mid i, j=1, \ldots, h\right\} \\
& \quad\cup\left\{12\left(b_{i j k k} \pm b_{i j j k} \pm b_{i i j k}\right) \mid i, j, k=1, \ldots, h\right\}\\
& \left.\quad\cup\left\{24 b_{i j k l} \mid i, j, k, l=1, \ldots, h\right\}\right)~.
\end{aligned}
\ee
We note that $\tilde{S}_2$ and $\tilde{S}_3$ `descend' from $\tilde{S}_1$ similarly to how $\tilde{S}_5$, $\tilde{S}_6$ and $\tilde{S}_7$ `descend' from $\tilde{S}_4$ and this is explained in detail in Appendix~\ref{app:basicGCDinvs}. This does by no means exhaust the list of GCD invariants. Two further examples, similar in form to $\tilde{S}_3$ and $\tilde{S}_7$, are given by:
\be
\label{eq:HubschStp3Stp7}
\begin{aligned}
\tilde{S}_3' &=\operatorname{GCD}\left(\left\{d_{i i i} \mid i=1, \ldots, h\right\}\right. \\
&  \cup\left\{3 d_{i i j} \mid i, j =1, \ldots, h\right\} \\
& \cup\left.\left\{6 d_{i j k} \mid i, j, k=1, \ldots, h\right\}\right)~,\\
\tilde{S}_7' & =\operatorname{GCD}\left(\left\{b_{i i i i} \mid i=1, \ldots, h\right\}\right. \\
&\quad\cup\left\{6 b_{i i j j} \mid i, j=1, \ldots, h\right\} \\
&\quad\cup\left\{4 b_{i i i j} \mid i, j=1, \ldots, h\right\} \\
& \quad\cup\left\{12b_{i j k k}\mid i, j, k=1, \ldots, h\right\}\\
& \left.\quad\cup\left\{24 b_{i j k l} \mid i, j, k, l=1, \ldots, h\right\}\right)~.
\end{aligned}\ee
Yet further examples can be obtained by using tensor products of $d_{ijk}$ and $c_{2,i}$ other than the completely symmetric one, $b_{ijkl}$. For example, a modification of $\tilde{S}_4$ leads to the invariant
\be
\label{eq:HubschSt4p}
\begin{aligned} 
\tilde{S}'_4&=\operatorname{GCD}\left(\left\{d_{ijk}c_{2,l} -d_{ijl}c_{2,k}\mid i, j, k,l=1, \ldots, h\right\}\right).
\end{aligned}\ee
Examples of this kind fit into a representation-theoretic approach, which we now discuss.

\subsection{Tensor powers and representation theory}\label{sec:TensorPowersAndInvariants}

The vector $c_{2,i}$ transforms in the fundamental $\mathbf{H}$ of $\operatorname{GL}(h, \mathbb{Z})$, and the tensor $d_{ijk}$ transforms in the representation $\mathbf{R} = \operatorname{Sym}^3(\mathbf{H})$. Each case leads to GCD invariants, but further such invariants can be obtained from tensor products of $\mathbf{R}$ and $\mathbf{H}$.

It turns out that symmetric tensor powers of $\mathbf{H}$ which correspond to polynomials in the integers $c_{2,i}$ do not lead to new invariants: the GCDs factorise and, as a result, there is only one independent GCD invariant for vectors in the fundamental (see Appendix \ref{app:basicGCDinvs}). On the other hand, symmetric tensor powers $\operatorname{Sym}^\delta(\mathbf{R})$, which correspond to polynomials in $d_{ijk}$, do lead to further GCD invariants (note that the entries of the intersection form fill out the full representation $\mathbf{R}$). 

Polynomials in both  $d_{ijk}$ and $c_{2,i}$ correspond to tensor products which involve both representations ${\bf H}$ and ${\bf R}$, such as $\operatorname{Sym}^\delta(\mathbf{R}\otimes \mathbf{H})$. In the simplest case of $\delta=1$, these correspond to the invariants $\tilde{S}_4$ and $\tilde{S}_4'$ above. In general, the analysis is more complicated, as expressions of the form $d_{ijk}c_{2,l}$ do not fill out the entire representation $\mathbf{R}\otimes \mathbf{H}$. As a result, it is not easy to confirm the existence of invariants, as many of the singlets actually evaluate to zero\footnote{For instance, for arbitrary vectors $v_i,w_j$ transforming in $\mathbf{R}$, $\det{v_iw_j}=0$ hence $v_iw_j$ do not fill out the representation $\mathbf{R}\otimes\mathbf{R}$.}, when expressed in terms of $c_{2,i}$ and $d_{ijk}$. Nonetheless, some nontrivial invariants can be identified.

We can analyse these invariants in detail using the inclusion of the discrete subgroup\footnote{The $\mathbb{Z}_2$ which extends the group to $\operatorname{GL}(h,\mathbb{Z})$ is `global' and so does not affect the Lie algebra representation structure, and indeed is therefore almost irrelevant for our purposes.} $\operatorname{SL}(h,\mathbb{Z})
\subset \operatorname{SL}(h,\mathbb{C})$. For illustration, we list the following branching rules (which can be obtained using computer algebra packages such as \texttt{LiE} \cite{LiE1992}) for the case $h=2$, $\mathbf{H}=\mathbf{2}$, $\mathbf{R} = \mathbf{4}$, focussing on the intersection numbers alone. The first few decompositions for polynomials up to degree $\delta=4$ are given by:
\be
\begin{aligned}
     \operatorname{Sym}^1(\mathbf{4})&=\mathbf{4}\\
     \operatorname{Sym}^2(\mathbf{4})&=\mathbf{3} \oplus\mathbf{7}\\
     \operatorname{Sym}^3(\mathbf{4})&=\mathbf{4} \oplus\mathbf{6} \oplus\mathbf{10}\\
     \operatorname{Sym}^4(\mathbf{4})&=\mathbf{1} \oplus\mathbf{5} \oplus\mathbf{7} \oplus\mathbf{9} \oplus\mathbf{13}~.
\end{aligned}
\label{eq:decompof2}
\ee

Each representation in these decompositions has an associated GCD invariant, which comes with a related, finite family of further GCD invariants derived in a manner analogous to how $\tilde{S}_2$ and $\tilde{S}_3$ and derived from $\tilde{S}_1$. The simplest of these new `polynomial GCD invariants' is discussed in Appendix \ref{app:basicGCDinvs}, along with the general algorithm for generating the higher degree cases.

We note that singlets in tensor representations, such as the singlet in the last row of Eq.~\eqref{eq:decompof2}, lead to `polynomial invariants' rather than polynomial GCD invariants (since the GCD of one number is simply the absolute value of this number). The representation theory will inform us where to look for these polynomial invariants, which, it turns out, can then always be written by appropriate contractions of $d_{i,jk}$, $c_{2,i}$ and the Levi-Civita tensor $\epsilon^{i_1\cdots i_h}$. We discuss such polynomial invariants and their construction in the following subsection.

\subsection{Polynomial invariants}
\label{sec:PolyInv}
The singlet in the fourth row of Eq.~\eqref{eq:decompof2} corresponds to the Cayley hyperdeterminant \cite{Cayley} of a trilinear form in two dimensions\footnote{The Cayley hyperdeterminant is a special case of a very general and powerful invariant, which we term the `Gelfand hyperdeterminant'~\cite{GelfandDiscrim}, related to the degeneracy of the multilinear form associated to a (not necessarily symmetric) array. This point is discussed in Appendix~\ref{app:polyinvs}, but we note that the degree of this polynomial is too high for any practical use beyond dimension 2.}. This can be written explicitly as $\Delta_{4,2}$, where the subscripts denote the degree and Picard number of the invariant:
\be 
\Delta_{4,2}(d_{ijk})= a^2 d^2 - 6 a b c d + 4 ( a c^3 +  b^3 d) -3 b^2 c^2 
\ee
where $d_{111}=a, d_{112}=b, d_{122}=c, d_{222}=d$.
In two dimensions, the ring of invariants of symmetric trilinear forms is generated by the Cayley hyperdeterminant\footnote{We note that the sign of the hyperdeterminant and the rank of the array viewed as a map $\operatorname{Sym}^2\mathbf{2}^* \rightarrow \mathbf{2}$ suffice to distinguish between symmetric three-arrays up to $\operatorname{GL}(2,\mathbb{R})$ equivalence. We can connect this classification to the complete classification of~\cite{BrodieGeodesicExtKahler} in terms of (class, rank): if the hyperdeterminant is zero, the (cl,rk)-type is $(1,1)$ or $(2,2)$. If the hyperdeterminant is positive, it is $(3,2)$, while if the hyperdeterminant is negative, the type is $(1,2)$. This is because the hyperdeterminant is the discriminant of the curve in $\mathbb{RP}^1$ induced by the cubic form associated to the trilinear form.}, but the situation is more complicated in higher dimensions.

In the three-dimensional case, there are two independent invariants of degrees four and six generating the ring\footnote{The ring includes the Gelfand hyperdeterminant which in this case has degree $36$.}. No method to predict the degree of all invariants is known, although it is relatively easy to establish some constraints---see Appendices \ref{app:polyinvs} and \ref{app:GeneratingPolWithLA} for details. We can, however, use the computer algebra package \texttt{LiE} to confirm exactly when they do occur.

The number, $N_{\rm gen}(h)$ of algebraically independent polynomial invariants (after taking into account possible syzygies) is bounded by $N_{\rm gen}(h)\leq N_{\rm upper}(h)$, where the number $N_{\rm upper}(h)$ is the na\"ive count for the number of basis-independent degrees of freedom given by
\be 
\label{eq:Ninv}
N_\text{upper}(h) =  \binom{h + 3 - 1}{3} - \left(h^2-1\right)~.
\ee
For the cases $h=1,2,3$, we have verified that this bound is saturated, that is $N_{\rm gen}(h)= N_{\rm upper}(h)$. For $h>3$, the number $N_{\rm gen}(h)$ is not known, but we expect---without strong evidence---that the bound continues to be saturated.

\begin{table}[h]
\begin{tabular}{|c||c|c|c|c|c|c|c|}
\hline
$h$   & ~1~ & ~2~ & 3   & 4    & ~5~  & 6     & 7  \\ \hline
$N_\text{upper}$ & 1 & 1 & 2   & 5    & 11 & 21    & 36 \\ \hline
(lowest) degrees & 1 & 4 & 4,6 & $8,16^{* \dagger}$ & 10 & 10*,12* & 14*\\ \hline
\end{tabular}
\caption{The maximum possible number of invariants and the lowest degrees of invariants of symmetric trilinear forms for $1\leq h\leq 7$. 
Starred degrees have not been evaluated for the KS data, and daggered degrees correspond to polynomials for which we have not constructed closed form expressions. Other invariants are discussed later, but have much higher degrees or involve the second Chern class.}
\label{tabdegnum}
\end{table}

In Table~\ref{tabdegnum}, we show the upper bound for the number of independent invariants from Eq.~\eqref{eq:Ninv}, and we list the degrees of the simplest invariants for each $h$, as determined with \texttt{LiE}.

For an $h$-dimensional trilinear form, there always exist an invariant polynomial with degree $2h$ (see Appendix~\ref{app:GeneratingPolWithLA} for more details), which can be seen to be generically nonvanishing. Such an invariant, in fact, be realised using the rarely-discussed `Pascal hyperdeterminant' ($\operatorname{PDET}$), defined as
\begin{align}
\operatorname{PDET} (a_{i_1,\dots,i_D})= & \frac{1}{D!} \sum_{\sigma_i \in S_D} \operatorname{sgn} \sigma_1 \cdot \operatorname{sgn} \sigma_2 \cdots \operatorname{sgn} \sigma_n \nonumber\\
& \times \prod_{i=1}^D a_{\sigma_1(i), \sigma_2(i), \ldots, \sigma_n(i)}\; ,
\label{pdet}
\end{align}
where $a_{i_1,\dots,i_n}$ is a tensor in $D$ dimensions. It is non-zero only for even tensor rank $n$ and it has polynomial degree $D$ much like the standard determinant. One is tempted to discard it, as we are mainly interested in three index tensors. However, we can consider the six index tensor obtained by symmetrising $d_{(ijk}d_{lmn)}$ and then evaluate the Pascal hyperdeterminant
\be\label{eq:PDET}
\operatorname{PDET}(d_{(ijk}d_{lmn)})~,
\ee
which is indeed of degree $2h$ in $d_{ijk}$ and it is non-vanishing\footnote{Note that using the anti-symmetrisation of $d_{(ijk}d_{lmn)}$ instead leads to a vanishing PDET.}. Unfortunately, the above formula for the PDET has exponential complexity and, for this reason, we use a recursive algorithm, presented in Ref.~\cite{AlgorithmForHdet}, to evaluate it\footnote{When evaluating expression~\eqref{eq:PDET}, it was useful to exploit the fact that the tensor $d_{(ijk}d_{lmn)}$ is symmetric. Memoisation also worked well. Incidentally, the fact that it can be written recursively explains the partial recurrence relation that we identified.}. This recursive algorithm has complexity $\mathcal{O}(2^{Dn}D^n) = \mathcal{O}(2^{6h}h^6)$, as $n=2\times 3$. 

\subsubsection*{Graphs and Levi-Civita tensors}
\label{sec:graphs}
\begin{figure}
    \centering
    \includegraphics[width=0.9\linewidth]{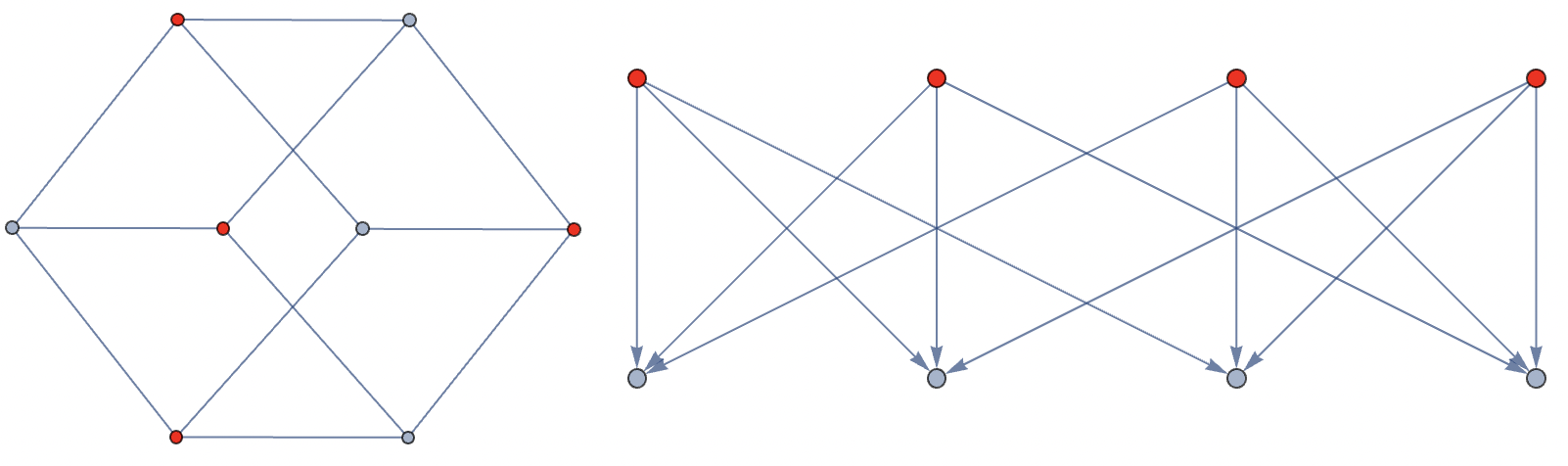}
    \caption{The bipartite graph denoting the index pattern for the $\delta=4$, $h=3$ `coincidental' invariant, firstly displayed undirected and then directed. Levi-Civita-type vertices are coloured red. Uniquely, for $h=3$ both the Levi-Civita and array vertices are trivalent.)}
    \label{fig:coincidentalgraph}
\end{figure}

We would now like to explain how some of the other invariants indicated in Table~\ref{tabdegnum}, in particular the quartic invariant for $h=3$ and the order $10$ invariant for $h=6$, arise. It is clear from Eq.~\eqref{pdet} that the Pascal hyperdeterminant can be written in terms of Levi-Civita symbols. This feature turns out to be more general, and it is a result from classical invariant theory---known as the first fundamental theorem of invariant theory---that \emph{all} polynomial invariants of $\operatorname{SL}(h,\mathbb{C})$ can be constructed in a similar fashion \cite{procesi2007lie}. We can therefore construct polynomial invariants by forming appropriate contractions between Levi-Civita symbols and the cubic form $d_{ijk}$. One immediate combinatorial requirement for a degree $\delta$ invariant to be non-vanishing is $\delta \leq \binom{3\delta/h}{3}$ and $3\delta/h$ integer. Such invariants can be represented as bipartite graphs with edges indicating the contractions between the $3\delta/h$ indices on Levi-Civita tensors and the $\delta$ indices on $d_{ijk}$ (for more details, see Appendix \ref{app:polyinvs}). We refer to these as contraction graphs. For example, the $h=3$, $\delta=4$ invariant,
\be \label{eq:i43}
I_{4,3}(d_{ijk})=\epsilon^{ilo}\epsilon^{jmr}\epsilon^{kps}\epsilon^{nqt} d_{ijk}d_{lmn}d_{opq}d_{rst},\ee
is represented by the contraction graph in Fig.~\ref{fig:coincidentalgraph}.
The rules for constructing these graphs have to be slightly modified if the second Chern class $c_{2,i}$ is included, in addition to $d_{ijk}$. Evidently, 
any given Levi-Civita symbol can only connect to one Chern class vector, or else the associated invariant vanishes. A simple method which guarantees a non-vanishing result is to start with an invariant associated to a degree $\delta$ invariant in $h-1$ dimensions for $d_{ijk}$ and then replace the Levi-Civita symbols $\epsilon^{i_1\cdots i_{h-1}}$ by $\eta^{i_1\cdots i_{h-1}}=\epsilon^{i_1\cdots i_{h}}c_{2,i_{h}}$.
This leads to an invariant of bi-degree $(\delta,\delta/(h-1))$ in the trilinear form and the second Chern class.
For example, to obtain an $h=4$ invariant which involves $d_{ijk}$ and $c_{2,i}$, we can start with the invariant $I_{4,3}$ for $d_{ijk}$ in Eq.~\eqref{eq:i43} and carry out the replacement $\eta^{ijk} = \epsilon^{ijkl}c_{2,l}$. The leads to the bi-degree $(4,4)$ invariant 
\be
I_{(4,4),4}(d_{ijk},c_l)=\eta^{ilo}\eta^{jmr}\eta^{kps}\eta^{nqt} d_{ijk}d_{lmn}d_{opq}d_{rst}
\ee
Using the same method, we can start with the Pascal hyperdeterminants~\eqref{eq:PDET} for dimension $h-1$ and obtain bi-degree $((2h-2),6)$ invariants in dimension $h$. Other ways of introducing the Chern class vector are discussed below. 

\subsubsection*{Determinant/rank/signature-type invariants}
\label{subseq:detrk}
Although all polynomial invariants can be written using contraction graphs, their evaluation quickly becomes unmanageable in large degree and/or dimension. We now discuss a subclass of polynomial invariants at large degree that can be evaluated using a different approach. This new approach allows us to introduce further new non-polynomial invariants corresponding to the rank and signature of certain linear maps. 

We begin the discussion by reconsidering the symmetric quadrilinear form $b_{ijkl}$, introduced in Section \ref{sec:GCD}, as a linear map $\hat{b}:~\operatorname{Sym}^2\mathbf{H}^*\rightarrow \operatorname{Sym}^2\mathbf{H}$. By Sylvester's law of inertia, we have a number of invariants: namely, these are the rank, signature, and determinant\footnote{This determinant was the only previously known polynomial invariant~\cite{hubschCYB}.} of the matrix~$\hat{b}$. Unfortunately, the rank becomes non-maximal when $h>3$, and consequently the determinant vanishes. More precisely, in the vast majority ($\gtrsim 99\%$) of cases considered in the list of KS CY manifolds (up to Picard number~$6$), the signature is zero and the rank is $2 h$. It follows that these provide very little discriminative power. 

However, we have identified another polynomial invariant, for $h$ even, also built from a linear map. This is much too large to be reasonably constructed using contraction graphs, since it has degree $h\binom{h+(h/2)-1}{h/2}$ which for $h=6$ is 336. Instead, we first introduce an auxiliary variable $z^i$ and consider the cubic polynomial $d_{ijk} z^i z^j z^k$. Taking the determinant of the Hessian, we are led to a degree $h$ polynomial given by
\begin{equation}
    P(z^i)=\det{d_{ijk}z^k}.
\end{equation}
By taking appropriate derivatives, we can form an $h$-array that is independent of $z^i$, as follows
\begin{equation}
\begin{gathered}
A_{i_1\dots...i_h}= A_{(i_1\dots...i_h)}=\frac{\partial^h P(z^i)}{\partial z^{i_1}\dots...\partial z^{i_h}}.
\end{gathered}
\label{eq:SymDet}
\end{equation}
Now, if $h$ is even, we can view $A_{i_1\dots...i_h}$ as a linear map $\hat{A}:~\operatorname{Sym}^{h/2}\mathbf{H}^*\rightarrow\operatorname{Sym}^{h/2}\mathbf{H}$. As was the case with $\hat{b}$, the determinant, rank, and signature of $\hat{A}$ are invariants. We call the polynomial invariant, coming from the determinant of $\hat{A}$, the `SymDet' of $d_{ijk}$. 

For a given $d_{ijk}$, computing $\operatorname{SymDet}(d_{ijk})$ has a computational complexity of $\mathcal{O}\left(\binom{h+(h/2)-1}{h/2}^3\right)$, given the second determinant is computed with Gaussian elimination. In practice, however, this is still more accessible than the Pascal hyperdeterminants (for small~$h$), precisely because it can be computed using Gaussian elimination. 

The SymDet invariant is frequently zero when applied to the KS CY data, presumably because there are many more zeros in an intersection form than there are in an arbitrary symmetric array. In any case, the rank and signature of $\hat{A}$ are still useful as, unlike $\hat{b}$, its rank and signature take many different values on the KS data.

Finally, there are similar invariants that vanish after some critical value of $h$ (for the same reason as the determinant of $\hat{b}$) and others that can be written down for $h$ satisfying other conditions. These are discussed in the Appendix~\ref{app:polyinvs}. One can also consider the rank of the intersection form, viewed as a map $\operatorname{Sym}^2\mathbf{H}^*\rightarrow \mathbf{H}$. However, this rank is maximal for all data used in this paper, and so provided no discriminative power.

\subsection{The discriminative power of the various invariants}
\label{sec:InvDiscrimPower}
Having introduced a large number of invariants, we now discuss their applicability and relative power to discriminate diffeomorphism types. 

Whilst polynomial invariants necessarily capture a considerable amount of information, any GCD invariants are less useful. If we take the GCD of $n$ numbers selected uniformly in some range $[1,N]$, a reasonably elementary argument yields that the probability that the GCD is $1$, in the limit of large $N$, is 1/$\zeta(n)$~\cite{DistribGCD}.
Using this $\zeta$ function argument as a rough approximation\footnote{\label{fn:footnoteignorerelations}We quotient by obvious relations like scaling; as such, any degree-$\delta$ polynomial invariant is only interesting (i.e. discriminating) if it is not unity after dividing by an appropriate factor $\operatorname{GCD}(d_{ijk})^\delta$. Similarly, there is an obvious relation between the GCD of $\mathbf{R}$ and $\Delta \otimes \mathbf{R}$, for $\Delta$ a singlet representation.}, it follows that the GCD invariants become less useful as the dimension of the representation increases. 

High-degree polynomial representations can still be relatively small in dimension. For example, in the $h=4$ case, there is a 4-dimensional irreducible representation in the branching of $\operatorname{Sym}^{11}\mathbf{R}$. This has a fair chance (approximately $1/\zeta(4)\approx10\% $) of having a non-trivial\footref{fn:footnoteignorerelations} GCD. Actually finding the explicit form of those polynomials is completely unfeasible, for the reasons mentioned above. Fortunately, all non-singlets have dimension greater than or equal to $h$, and so for higher $h$ even these `coincidental' low-dimensional irreducible representations should become less powerful. However, we note that the zeta-function calculation is an exceedingly poor approximation, as typical values in the data are far from being uniformly distributed.

To exemplify how GCD invariants perform, we consider the Picard number $h=6$ case. There are 128 distinct Hodge numbers, and GCD invariants suffice to delineate 2092 diffeomorphism classes. All non-GCD invariants delineate $51,330$ classes. All invariants combined, increase this lower bound to $52,361$.

We note that the GCDs are the only invariants used here which guard against $\operatorname{SL}(h,\mathbb{Q})$ relations. We do not expect them to provide a total classification. There may well exist other invariants, such as the analogues of the `spinor genus' for quadratic forms, which would provide more effective protection. These are unfortunately beyond the scope of this paper, but we note that, for example, it is possible that the techniques developed for quadratic forms could be applied to either of the linear maps $\hat{b}$ or $\hat{A}$ created in the previous subsection.

At each level $h$, then, there is a finite number of polynomial invariants, as well as an infinite number of GCD invariants. All of these are, in general, difficult to construct. None have particularly favourable complexity for evaluation. Computationally, the first (lowest degree) invariants are reasonably tractable\footnote{The $2h$ invariants were not actually evaluated for the FRST list at level $h=6$, as they take $\mathcal{O}$(10s) to evaluate for each triangulation. Instead, the ($2h-2$)-invariants were used.} for individual cases up to $h\simeq 8$. Due to the large number of manifolds descending from the KS list, we avoid evaluating invariants that take~$\gtrsim 5s$ to compute for $h=6$.

\subsection{Real equivalence of manifold data}

While not of relevance for the problem at hand, we briefly discuss what the situation for invariants would be if we were interested in $\operatorname{GL}(h,\mathbb R)$ equivalence of manifold data. For some other purposes, the real equivalence of the intersection form can be of use\cite{BrodieGeodesicExtKahler}. In this case, GCD invariants are no longer useful. Polynomial invariants identified become \emph{relative} invariants, transforming with a suitable power of the determinant of the transformation matrix. If this is an even power, the sign of the relative invariant is an invariant. 

Furthermore, a suitable ratio of two different relative invariants becomes itself invariant. For example, for the Picard number $3$ singlets under the special linear group (the quartic and sextic invariants $I_{4,3}$ and $I_{6,3}$), the ratio $I_{4,3}^3/I_{6,3}^2$ is an invariant. This now represents the only remaining continuous and basis-independent degree of freedom. As before, ranks and signatures remain invariant.

\section{Direct identification of basis transformations}\label{sec:search_alg}
As before, we consider two smooth simply connected CY threefolds $X$ and $X'$ with torsion-free cohomology. Recall that we have introduced bases $(D_i)$, where $i=1,\ldots ,h$, of $H^2(X,\mathbb{Z})$ and $(D_i')$, where $i=1,\ldots ,h'$, of $H^2(X',\mathbb{Z})$ and the intersection forms and second Chern classes, relative to these bases, are denoted by $(d_{ijk},c_{2,i})$ and $(d'_{ijk},c_{2,i}')$, respectively. If $h\neq h'$ or if any of the invariants introduced in the previous section differ, the manifolds $X$ and $X'$ are clearly not diffeomorphic, so let us instead assume $h=h'$ and identical values for all invariants. In this case, we have to decide whether a basis transformation $P\in \operatorname{GL}(h,\mathbb{Z})$ which satisfies Eqs.~\eqref{eqns_for_P} exists. Clearly, this is difficult (except, possibly, for small values of $h$) so instead we ask the following related and simpler question.\\[2mm]
{\bfseries Problem:} {\itshape For a given $k_{\rm max}>0$, is there a $P\in \operatorname{GL}(h,\mathbb Z)$ with all $|{P_i}^r|\leq k_{\rm max}$  that satisfies Eqs.~\eqref{eqns_for_P}?}\\[2mm]
In this section, we describe an algorithm for solving this simplified problem. In Section~\ref{sec:Results} we will employ this algorithm to obtain upper bounds on the numbers of diffeomorphism classes in the Kreuzer-Skarke list for all cases with $h\leq 6$, by removing equivalences within the classes of manifolds with the same invariants. 

The algorithm in question relies
on the observation that  $H^2(X,\mathbb Z)$ is also the group of isomorphism classes of holomorphic line bundles over $X$ (with the tensor product of line bundles as the group operation).
We denote the line bundle $L$ with first Chern class $c_1(L)=k^iD_i$ by ${\cal O}_X(k)$. The Hirzebruch-Riemann-Roch theorem states that its index can be computed from
\begin{equation}\label{index}
\begin{aligned}
\chi(X,L) &=  \frac{1}{12} \left( 2\,c_1(L)^3 + c_1(L)\,c_2(TX)\right) \\
&=   \frac{1}{6} d_{ijl}k^ik^jk^l + \frac{1}{12} k^i c_{2,i}\; .
\end{aligned}
\end{equation}
Suppose two manifolds $X$ and $X'$ are diffeomorphic via a matrix $P\in \operatorname{GL}(h,\mathbb{Z})$ which satisfies Eqs.~\eqref{eqns_for_P}. Then, for a line bundle $L={\cal O}_X(k)$ there exists a unique line bundle $L'={\cal O}_{X'}(k')$ over $X'$, given by $k'={P^{-1}}^Tk$, such that
\begin{equation}\label{eq:chiLLprime}
\chi(X,L) = \chi(X',L')~.
\end{equation}
Moreover, for related line bundles, $k'={P^{-1}}^Tk$, it is clear that the two terms in Eq.~\eqref{index} must be invariant separately, so that, defining
\begin{equation}\label{eq:lbconds}
\begin{array}{rclcrcl}
\chi_d(k) &=&d_{ijl}k^ik^jk^l &&\chi'_d(k')&=&  d'_{ijl}{k'}^i{k'}^j{k'}^l\\
 \chi_c(k) &=&k^ic_{2,i}&&\chi'_c(k')&=& {k'}^ic'_{2,i}\; ,
\end{array}
\end{equation}
we must have $\chi_d(k)=\chi'_d(k')$ and $\chi_c(k)=\chi'_c(k')$.
These conditions are stronger than Eq.~\eqref{eq:chiLLprime} and, in fact, imply that
\begin{equation}
\chi(X,{\cal O}_X(nk)) = \chi(X',{\cal O}_{X'}(nk'))
\end{equation}
for all $n\geq 1$. Moreover, if we have line bundles $L_a={\cal O}_X(k_a)$, where $a=1,\ldots ,m$, and the corresponding line bundles $L_a'={\cal O}_{X'}(k_a')$, related under the diffeomorphism (so that $k_a'={P^{-1}}^Tk_a$), the linear combinations
\begin{equation}
 L={\cal O}_X(n^ak_a)\;,\quad
 L'={\cal O}_{X'}(n^ak_a')\; ,
\end{equation}
where $n^a\in\mathbb{Z}$, are related under the diffeomorphism as well.

The basic idea of the algorithm, described below, is that the invariance of $\chi_d$ and $\chi_c$ in Eq.~\eqref{eq:lbconds} (and linear combinations preserving this invariance) places strong constraints on viable basis transformations $P\in \operatorname{GL}(h,\mathbb{Z})$.

\subsection{The algorithm}\label{sec:alg}
We are now ready to present the algorithm which solves the problem stated above, that is, which decides whether a basis transformation $P\in \operatorname{GL}(h,\mathbb Z)$ with entries bounded by $|{P_i}^r|\leq k_{\rm max}$ exists.\\[2mm]
{\bfseries Step 1:} For a `basis' of line bundles ${\cal O}_X(k_i)$, where $i=1,\ldots ,h$ and $c_1({\cal O}_X(k_i))=D_i$, evaluate the 
quantities $\chi_d(k_i)$ and $\chi_c(k_i)$ from Eq.~\eqref{eq:lbconds}.\\[2mm]
{\bfseries Step 2:} Evaluate 
$\chi_d(k')$ and $\chi_c(k')$ for all line bundles ${\cal O}_{X'}(k')$ with $|{k'}^i|\leq k_{\rm max}$. Then, for each $i=1,\ldots ,h$, select for such $k'$ the subset $S_i=\{k'\,|\, \chi'_d(k')=\chi_d(k_i)\mbox{ and }\chi'_c(k')=\chi_c(k_i)\}$.\\[2mm]
{\bfseries Step 3:} Construct the set $S_{1,2}$ by selecting pairs $(k'_1,k'_2)\in S_1\times S_2$ such that $\chi_d(k)=\chi'_d(k')$ and $\chi_c(k)=\chi'_c(k')$ for any linear combinations $k=n_1 k_1 + n_2 k_2$ and $k'=n_1 k'_1 + n_2 k'_2$. In practice, checking for a single pair of (sufficiently large) integers $(n_1,n_2)$ is enough.\\[2mm] 
{\bfseries Step 4:} If $S_{1,2}$ is non-empty, obtain the set $S_{1,2,3}$ by the analogous construction. If $S_{1,2,3}$ is non-empty, proceed to constructing $S_{1,2,3,4}$ and so on.\\[2mm] 
{\bfseries Step 5:} If repeating Step 4 proceeds to the construction of a non-empty set $S=S_{1,2,\ldots,h}$, then $S$ contains a number of potential integral basis transformations $P=(k'_1,\ldots ,k'_h)$. For each such $P$, check if Eqs.~\eqref{eqns_for_P} hold. In general, if a basis transformation is found, it is not unique.\\[2mm] 
If $h=1$, Steps 3 and 4 are omitted and if $h=2$, Step 4 is omitted.
The slowest part of the above algorithm is Step 2. It involves a search over all line bundles ${\cal O}_{X'}(k')$ within the search box $|{k'}^i|\leq k_{\rm max}$ which satisfy the Diophantine equations $\chi'_d(k')=\chi_d(k_i)$ and $\chi'_c(k')=\chi_c(k_i)$ for the given basis ${\cal O}_X(k_i)$ of line bundles on $X$. The large computational cost is due to the number $(2k_{\rm max}+1)^{h}$ of line bundles within this search box. For large values of $h$, this becomes unfeasible even for small values of~$k_{\rm max}$.

\subsection{Other approaches and extensions}
\begin{figure}[!ht]
\centering
\resizebox{0.5 \textwidth}{!}{%
\begin{circuitikz}
\tikzstyle{every node}=[font=\small]
\node [font=\large] at (2,8.99) {Input};
\draw [ -Stealth] (2.75,9)--(4.75,10.5);
\draw [ -Stealth] (2.75,9)--(4.75,7.5);
\node [font=\small] at (3.5,10) {$h$};
\node [font=\small] at (3.5,8) {$h$};
\draw  (4.75,8) rectangle  node {\large ${(P^{-1})}^T$} (6.25,6.75);
\draw  (7.25,8) rectangle  node {$[\chi'_d(k'),\chi'_c(k')]$} (9.5,6.75);
\draw [ -Stealth] (6.25,7.4)--(7.25,7.4);
\draw  (4.75,11) rectangle  node {$[\chi_d(k),\chi_c(k)]$} (7,9.75);
\draw  (10.5,10) rectangle  node {\large $L$} (12,8.75);
\draw [ -Stealth] (12,9.5)--(13.75,9.5);
\draw [](7,10.5) to[short] (8.5,10.5);
\draw [](8.5,10.5) to[short] (8.5,9.75);
\draw [](9.5,7.25) to[short] (10,7.25);
\draw [](10,7.25) to[short] (10,9);
\draw [ -Stealth] (8.5,9.75)--(10.5,9.75);
\draw [ -Stealth] (10,9)--(10.5,9);
\node [font=\small] at (8,10.75) {2};
\node [font=\small] at (9.75,8.25) {2};
\node [font=\small] at (6.8,7.75) {$k'$};
\node [font=\small] at (14,9.5) {1};
\end{circuitikz}
}
\caption{Neural network that learns the required $\operatorname{GL}(h,\mathbb Z)$-transformation $P$.}\label{fig:NN}
\end{figure}
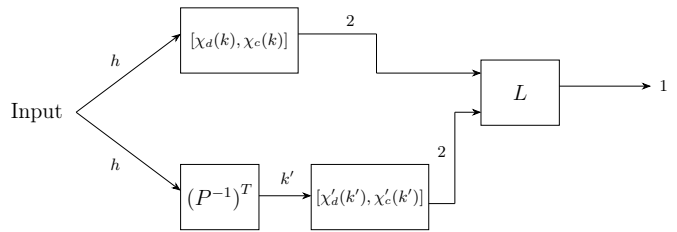

Since the above `classical' algorithm becomes too slow for large values of $h$, it is worth trading completeness for speed by employing heuristic methods. Such methods, including genetic algorithms, reinforcement learning and quantum annealing have been used to solve similar Diophantine equations in Refs.~\cite{Halverson:2019tkf, Cole:2019enn, Cole:2021nnt, Constantin:2021for, Krippendorf:2021uxu, Abel:2021rrj, Abel:2021ddu, Abel:2022wnt, Abel:2023zwg, Berglund:2023ztk}. These results show that a good fraction of the solutions can be found by checking a tiny sample of the search space, provided a non-trivial number of solutions exists. The fundamental feature of heuristic searches is to rank the possible alternatives at each branching step based on the available information in order to decide which branch to follow. This approach can considerably speed up the identification of solutions. However, in case of an empty search result, the absence of solutions is by no means guaranteed but only established with a certain level of confidence. In the following, we briefly discuss the various heuristic methods we have considered in turn.\\[2mm]
{\bfseries Newton-Raphson minimisation.} Following Ref.~\cite{CandelasHowManyCICYs}, direct Newton-Raphson minimisation of a `loss' which measures the failure to satisfy Eqs.~\eqref{eqns_for_P}
is reasonably successful in finding rational transformations $P\in \operatorname{GL}(h,\mathbb{Q})$. The method involves the computation of the inverse of a ${h}^2$-dimensional matrix, which is the main limiting factor. We did not pursue this method further as the result is stochastic (dependent on initialisation), and it produces matrices in $\operatorname{GL}(h,\mathbb{Q})$, rather than the required ones in $\operatorname{GL}(h,\mathbb{Z})$.\\[2mm]
{\bfseries Genetic algorithms.} It is possible to search for $\operatorname{GL}(h,\mathbb{Z})$ matrices satisfying Eqs.~\eqref{eqns_for_P} using a genetic algorithm. Given the intersection numbers and second Chern classes of two manifolds, the environment consists of $h\times h$ integer matrices $P$ with entries restricted as $k_{\rm min}\leq {P_i}^r\leq k_{\rm min}+2^{n_{\rm bits}}-1$. Then every entry of $P$ can be represented by $n_{\rm bits}$ bits and the entire matrix by a bit list of length $h^2n_{\rm bits}$. The fitness function needs to measure the failure to satisfy Eqs.~\eqref{eqns_for_P} as well as incorporate the condition ${\rm det}(P)=\pm 1$ and an obvious choice is
\begin{equation}
\label{fitnessfunction}
\begin{aligned}
    f(P) &=- w_1 \left | d_{ijk} -  P_i^rP_j^sP_k^t d'_{rst} \right | \\ 
&  - w_2 \left |c_{2,i} - P_i^r c'_{2,r} \right |   - w_3 \lvert\,  |\!\det(P)|  - 1\rvert~.
\end{aligned}
\end{equation}
Here, $w_1, w_2$ and $w_3$ are positive weights that can be adjusted to optimise performance. We have implemented and extensively tested such an algorithm but, it turns out, while its performance is much better than random search it is inferior to the algorithm described in Section~\ref{sec:alg}.\\[2mm]
{\bfseries Neural nets.} Another approach is to train a neural network to learn $\operatorname{GL}(h,\mathbb Z)$-transformations. We present a possible architecture in \fref{fig:NN}. The input consists of vectors $k\in\mathbb{Z}^k$ which described line bundles~${\cal O}_X(k)$. The upper branch of the network computes the invariants $\chi_d(k)$ and $\chi_c(k)$ for the manifold $X$ from Eq.~\eqref{eq:lbconds}. The lower branch first transforms to $k'=(P^{-1})^Tk$, where the matrix $(P^{-1})^T$ represent the weights of the layer, before computing the invariants $\chi'_d(k')$ and $\chi'_c(k')$ for the manifold $X'$. The final loss layer measures the difference between these invariants, that is, $L=(\chi_d'(k')-\chi_d(k))^2+(\chi'_c(k')-\chi_c(k))^2$. The idea is that, after training this network with a training set $\{k\}$ of integer vectors, the weights in $(P^{-1})^T$ have settled to a viable transformation matrix. This does indeed work for many cases, but it is, unfortunately, a very time-consuming process, as the network has to be trained for every pair of manifolds. Of course, it naturally leads to real rather than integer transformation matrices.

\noindent {\bfseries Image Recognition.} 
Another approach we have investigated is inspired by image processing techniques. Point-set registration is the process of aligning two point sets which are misaligned by some spatial transformation. Working again with line bundle vectors, we consider two copies of $\mathbb{Z}^{h}\subset\mathbb{R}^{h}$ and `colour' the line bundle vectors $k$ of the first set with the values $(\chi_d(k),\chi_c(k))$ from Eq.~\eqref{eq:lbconds}, and the line bundles vectors $k'$ of the second set with $(\chi'_d(k'),\chi'_c(k'))$. The result is two `coloured' copies of (some finite subset of) $\mathbb{Z}^{h}$, and we wish to find a point-set registration of these copies consistent with the colouring.
We have used a coherent point drift (CPD) point-set registration algorithm \cite{MyronenkoSong}, which maximises the likelihood of a Gaussian mixture model via an expectation-maximisation algorithm. We modified the code in Ref.~\cite{pycpd} to use colours. In the language of Ref.~\cite{MyronenkoSong}, we have adjusted the code to block-diagonalise the alignment probability matrix, and restrict the transformation matrix to be in $\operatorname{GL}(h,\mathbb{R})$. The `noise' parameter, which works by adding a uniform distribution to the Gaussian mixture model, partially accounts for the fact that the finite subsets of the two copies of $\mathbb{Z}^h$ are not mapped precisely to each other under the $\operatorname{GL}(h,\mathbb{Z})$ transformation. Tuning of the noise parameter is required, and it is a good idea to restrict to a small number of colours. Somewhat surprisingly, this algorithm performs reasonably well, despite being heuristic: for a given box size, it would typically identify the same transformation matrix as the systematic search algorithm of Section~\ref{sec:alg}. However, whilst the latter is more memory-intensive, the former is entirely deterministic. For this reason, we did not pursue the image recognition approach further.

\begin{widetext}$~~~~~~~$
\begin{table*}[htb]
  \centering
    \begin{tabular}{|c||c|c|c|c||c|c|c|}
    \hline $h$ &\# Polytopes & \# FRSTs &  \# Distinct FRSTs & Hodge \#s  & Lower Bound & Upper Bound &$k_{\text{max}}$\\
    \hline 1 & 4&4 & 4 & 4& 4 (4) & 4 &$\infty$\\ 
    \hline 2 & 34& 46 & 36 &16& 25 (27) & 27 &$\infty$ \\ 
    \hline 3 & 238 &517 & 291 &41& 171 (183) &  183&$\infty$ \\ 
    \hline 4 & 1,179 &5,324 & 1,948 &86& 1,113 & 1,183 & 15  \\
    \hline 5 & 4,897& 56,714 & 13,330 &113& 7,630 &  8,023 & 12 \\ 
    \hline 6 & 16,608 &584,281 & 83,906 &128&  52,361 & 54,939  &5 \\ \hline
  \end{tabular}
  \caption{Upper and lower bounds on the number of diffeomorphism classes present in the KS list of simply connected CY threefolds up to Picard number~$6$. 
  Lower bounds have been obtained using the invariants summarised in Table~\ref{tab:mergedInvars}. Upper bounds have been obtained by explicitly identifying basis transformations, using constraints related to the index of line bundles. The bracketed lower bounds denote the exact numbers of classes decided by symbolic solutions in Mathematica.
  The fourth column indicates the number of FRSTs with trivially distinct topological data. The fifth column indicates the lower bounds given by the Hodge numbers alone. The last column indicates the range of line bundles used in the derivation of the upper bounds. }
  \label{tab:Results}
\end{table*}
\end{widetext}
\section{Application to the Kreuzer-Skarke List}\label{sec:Results}
The largest known list of CY threefolds descends from the KS list \cite{kreuzer2000complete}, where the manifolds are represented by hypersurfaces in toric varieties corresponding to reflexive polytopes enumerated by Kreuzer and Skarke. Each of the 473,800,776 reflexive polytopes can be used to generate potentially many topologically inequivalent CY manifolds by triangulation, which corresponds to a choice of particular hypersurface. 

Distinct triangulations of the same polytope can correspond to distinct manifolds, but they must have the same Hodge numbers as that property descends from the polytope. Any two triangulations of different polytopes might be diffeomorphic, but manifolds from the same polytope often seem to have numerically similar or even identical data and so are particularly likely to be equivalent. We use \texttt{cytools}~\cite{cytools} to generate all manifolds corresponding to favourable\footnote{For a definition of favourability, see \cite{Demirtas:2018akl}. At least at low $h$, most polytopes in the list are favourable. For $h=1,2,3,4,5,6$, the numbers of non-favourable polytopes are $0, 0, 1, 12, 93, 493$. Of \emph{all} favourable polytopes, a further 14 give rise to manifolds which are not simply connected. All 14 have $h^{1,1}\in[1,4]$.} polytopes for $h\leq6$, and compute their corresponding data. Note that many of these triangulations have exactly the same data and are therefore clearly isomorphic. It suffices, then, to remove exactly duplicate data, and thereby state the na\"ive upper bound on the number of manifolds: we call this the number of \emph{trivially distinct} FRSTs. The results of this section are summarised in Table~\ref{tab:Results} and in Figs.~\ref{fig:TriangEquivPlot} and \ref{fig:NumberTriangdivNumberPolytopes}.

\subsection{Evaluating invariants}

For each value of $h$, we evaluate some of the invariants identified in Section~\ref{sec:LowerBound}; precisely which invariants were used is explained in Table~\ref{tab:mergedInvars} and Appendix~\ref{app:data}.

We note that the polynomial invariants become increasingly difficult to compute with larger $h$. Given the exponential increase in the number of FRSTs at each Picard number, we only computed invariants for $h\leq 6$. If we do not compute sufficiently powerful invariants, some equal-invariant classes become impractically large, and we cannot reasonably run a pairwise-comparison upper bound algorithm. Consequently, in this work, we present only the bounds for $h \leq 6$. 

The lower bounds, provided by these invariants, are presented in Table~\ref{tab:Results}.

\subsection{Running the systematic search algorithm}

Within each class, we use a union-find method to identify equivalent manifolds, searching for a $\operatorname{GL}(h, \mathbb Z)$ transformations using the algorithm described in Section~\ref{sec:alg}. For all $h\leq6$, we iteratively apply the algorithm at increasing values of $k_\text{max} = 1,2,3,\dots$, up to the maximum given in the last column of Table~\ref{tab:Results}. The systematic search algorithm was compiled to the Wolfram virtual machine, and the rest of the program was realised in the Wolfram language in Mathematica, and run on a laptop for $h=1,2,3,4$ and an HPC cluster for $h=5,6$. For $h>5$, we found it advisable to set a constraint of $\sim 1$GB to memory usage (in any given instance of the search algorithm), as in rare cases the sets $S_{1,2,\dots ,k}$ (for $k<h$) become extremely large\footnote{These typically corresponded to cases where many of the integers $d_{ijk}$ and $c_{2,r}$ vanished.}. 

\begin{figure*}
    \centering
    \includegraphics[width=14cm]{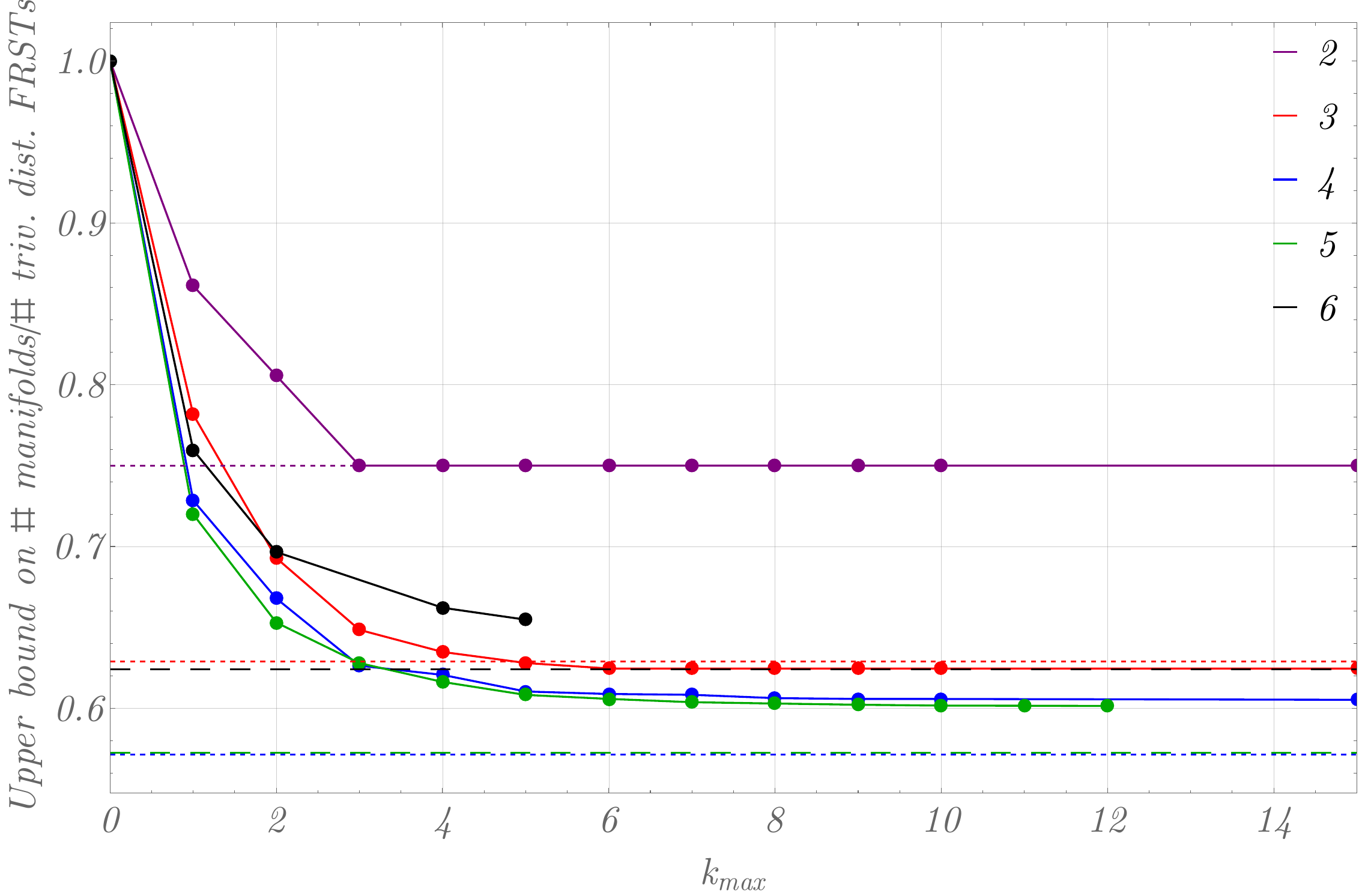}
    \caption{The solid lines are upper bounds on the number of diffeomorphism classes per distinct triangulation, for $h \leq 6$, identified by the algorithm introduced Section~\ref{sec:alg} as $k_{\text{max}}$ is increased. For $h=2,3,4,5,6$, $k_{\text{max}}=\infty,\infty,15,12,5$. Lower bounds for each value of $h$ are indicated with dashed lines in the appropriate colour.}
    \label{fig:Saturation}
\end{figure*}

The evolution of the upper bound on the number of diffeomorphism classes, upon changing $k_\text{max}$, is plotted in Figure~\ref{fig:Saturation}. We stop increasing $k_\text{max}$ after either observing saturation, or after runtime exceeded 48 CPU hours. We see that saturation occurs for $k_{\text{max}}\sim5$, providing good evidence that most manifold equivalences have been identified in this search. The final result for each $h$ is given in Table~\ref{tab:Results}.

Looking at the resulting data, the nature of the equivalence classes, resulting in the KS list, appears difficult to predict. For example, there were exactly 648 incidences of one particular manifold in Picard number $h=6$, from 17 different polytopes. 627 of them have numerically distinct data, but all transition matrices between those 627 were found after searching for basis transformations with entries of up to $5$. By contrast, another manifold is realised 373 times with precisely the same numerical data.

\subsection{Exact determination of the number of manifolds for $h\leq 3$, and rational equivalence}

For $h\leq 3$, Mathematica's symbolic solver is able to decide equivalence by directly solving Eq. \ref{eqns_for_P}. It follows that, for those cases, we should be able to determine the true number of diffeomorphism classes. We do this by running the union-find algorithm coupled to Mathematica's $\texttt{Reduce}$ function, only demanding that the Hodge numbers should be equal. At Picard numbers $h=1,2,$ and $3$, we find that there are exactly 4, 27, and 183 classes of simply connected manifolds with data related under $\operatorname{GL}(h,\mathbb{Z})$.

How close can we get to the exact number of $183$ diffeomorphism classes for $h=3$, using the invariants from Section \ref{sec:LowerBound}? Polynomial invariants and the Hodge numbers only discriminate 150 diffeomorphism types. In fact, $I_{4,3}, I_{6,3}$ (from Section \ref{sec:graphs}) and $\operatorname{det}\hat{b}$ (from Section \ref{subseq:detrk}) suffice to achieve this lower bound. By using GCD invariants, we are able to increase the bound from 150 to 171. However, our invariants are insufficient to saturate the lower bound at 183.  

Interestingly, if we allow $\operatorname{GL}(h,\mathbb{Q})$ equivalence, there are precisely 4, 22, and 150 distinct sets of topological data respectively, and all the $\operatorname{GL}(h,\mathbb{Q})$ transformations have determinant $\pm 1$. Somewhat surprisingly, this means that the polynomial invariants are still preserved---even though rational transformations need not have determinant $\pm 1$: we should really only trust relative invariants, which exist for $h=3$ as $I_4^3/I_6^2$, and do not exist for $h=1$ and $2$. This explains the $150$ topological types discriminated by polynomial invariants.

\subsection{Extrapolation to the entire KS list}
\begin{figure}
    \centering
    \includegraphics[width=\linewidth]{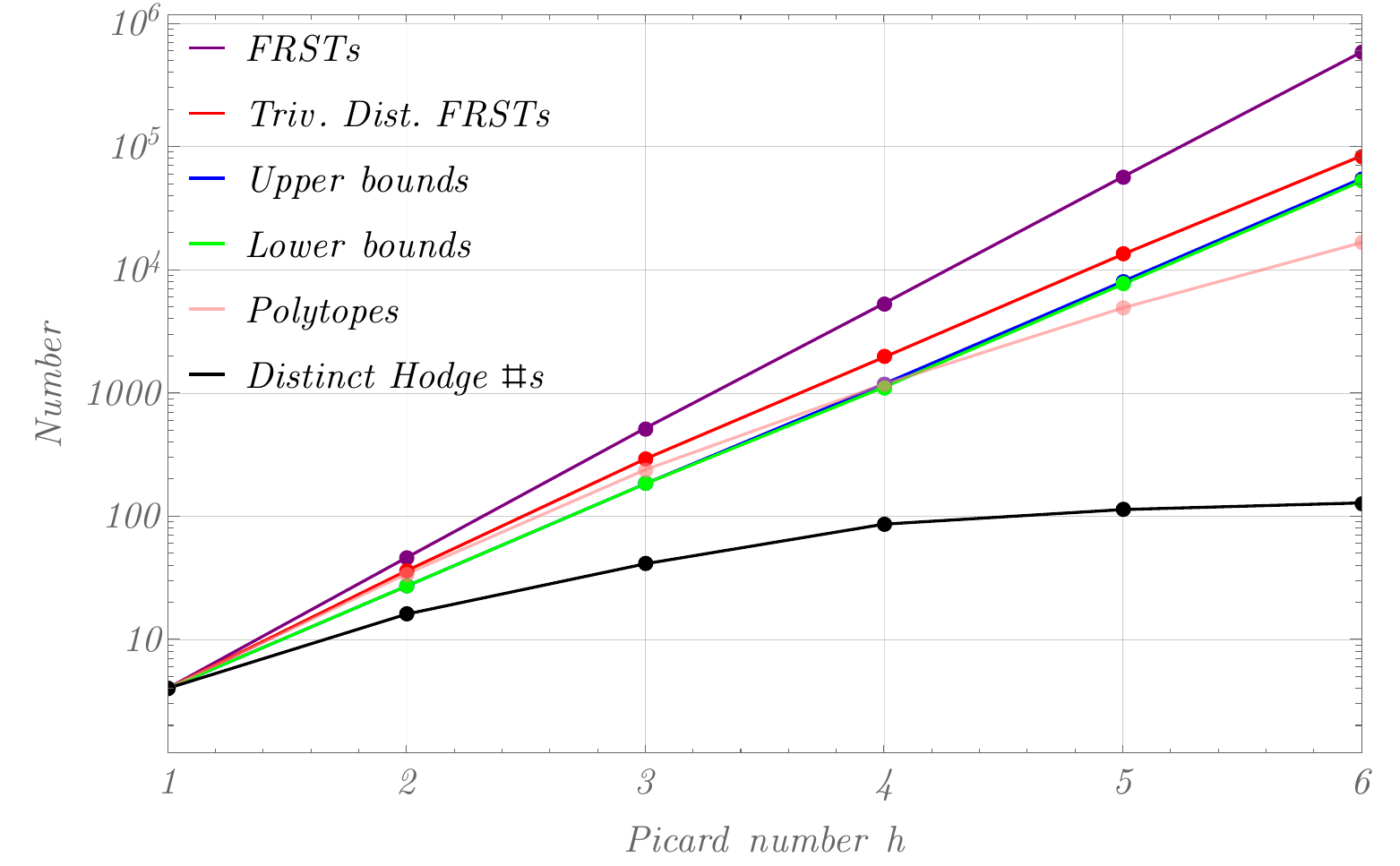}
    \caption{Plot corresponding to Table~\ref{tab:Results} on a logarithmic scale. Here, we plot the lower (from the invariants) and upper (from the systematic search) bounds on the number of distinct manifolds. We also plot the bare numbers of polytopes and triangulations, as well as the number of numerically distinct triangulations (i.e. triangulations with exactly the same topological data).}
    \label{fig:TriangEquivPlot}
\end{figure}
\begin{figure}
    \centering
    \includegraphics[width=\linewidth]{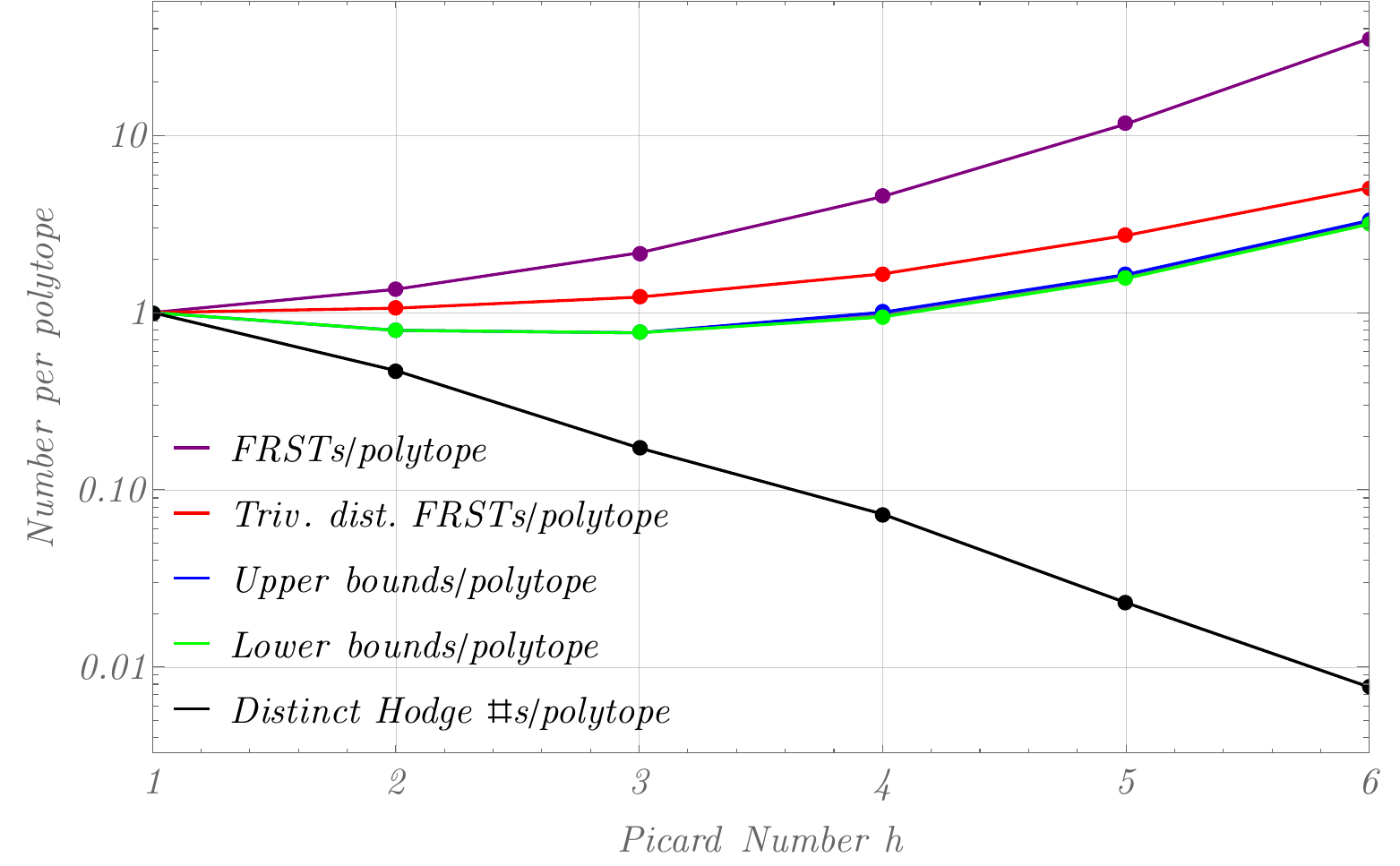}
    \caption{Average number of distinct manifolds per polytope from Table~\ref{tab:Results}.}
    \label{fig:NumberTriangdivNumberPolytopes}
\end{figure}
Looking at Figs.~\ref{fig:TriangEquivPlot} and \ref{fig:NumberTriangdivNumberPolytopes} the number of FRSTs, as well as the number of diffeomorphism classes, appears to grow exponentially with $h$. An exponential fit for $h\leq6$ gives the number of FRSTs as Eq.~\eqref{eq:ntr} and the number of diffeomorphism classes as Eq.~\eqref{eq:ndiff}. We therefore predict $(6.023\pm0.002)\times10^6$ FRSTs, with $(3.3\pm0.4)\times 10^5$ diffeomorphism classes at Picard number $h=7$. By way of comparison, using $\texttt{cytools}$, we find 5,990,333 favourable FRSTs for $h=7$, which is very close to our estimate. Of these, 522,388 are numerically distinct, which is compatible with our prediction of the number of diffeomorphism classes. 

A na\"ive extrapolation of the above fits to the entire (favourable) KS list, leads to $(1.6\pm0.1)\times10^{497}$ FRSTs and to between $10^{396}$ and $10^{401}$ diffeomorphism classes. As might be expected, these estimates are dominated by the single polytope which leads to manifolds $X$ with $h^{1,1}(X)=491$.

\subsection{CY manifold class data}
The list of CY manifolds from the KS list, up to $h=6$, along with their evaluated invariants can be found on the site specified in Appendix \ref{app:data}. These data files also include the explicit maps constructed between different manifolds with equal invariants when available.

Two files are provided for each Picard number $h=h^{1,1}(X)$. The first is a list of the manifolds and invariants, and is contained in the file $\texttt{ManifoldData.zip}$. The second is a list of equivalences, contained in the file $\texttt{EquivalenceData.zip}$. The equivalence data is designed to be handled in Mathematica, and comprises three parts. Firstly, we have a list of classes of FRSTs which all have the same invariants and are not linked to each other by $\operatorname{GL}(h,\mathbb{Z})$ transformation. Secondly, we have an association taking a pair of manifolds to a list of $\operatorname{GL}(h,\mathbb{Z})$ transformation matrices which realise the transformation between the two. Thirdly, we have an association describing the FRSTs with numerically duplicate data, taking a single manifold in the first or second list to a list of other manifolds. For a full explanation, see Appendix \ref{app:data}.

Finally, in $\texttt{EquivalenceData.zip}$, we also include two files which contain the $GL(h,\mathbb{Q})$ relations between manifold data for $h=2,3$. These files use exactly the same formatting as the other equivalence data.
\section{Conclusions}\label{sec:Conclusions}

In this work, we have placed bounds on the number of diffeomorphism classes present in the list of CY three-folds of low Picard number derived from the KS list. Wall's theorem asserts that two CY three-folds are diffeomorphic iff their Hodge numbers, intersection forms and second Chern classes coincide. In practice, the intersection form and the second Chern class are specified relative to an integral basis on $H^2(X,\mathbb{Z})$, so checking equality involves finding suitable $\operatorname{GL}(h,\mathbb Z)$ basis transformations, where $h=h^{1,1}(X)$. This can be difficult, particularly for larger $h$, and, as a result, it is not \emph{a priori} clear how many inequivalent CY three-folds are contained in the dataset. 

We have tackled this problem by using sets of complementary methods: (i) by using $\operatorname{GL}(h,\mathbb{Z})$ invariants constructed from the intersection numbers and the components of the Chern class and (ii) by using an algorithm, based on line bundle invariants, to find suitable $\operatorname{GL}(h,\mathbb{Z})$ transformations and (iii) by directly solving, for low Picard numbers, $h\leq 3$, the relevant Eqs.~\eqref{eqns_for_P} for the transformation matrix $P\in \operatorname{GL}(h,\mathbb{Z})$ using computer algebra methods.

For the first method, we have relied on known $\operatorname{GL}(h,\mathbb{Z})$ invariants from the literature~\cite{hubschCYB} as well as on novel invariants constructed here for the first time. These invariants, discussed in detail in Section~\ref{sec:LowerBound}, split manifolds with the same Hodge numbers into subclasses, thereby providing lower bounds on the number of diffeomorphism classes in the KS list. For higher Picard number, some of these invariants, specifically the polynomial ones, become too computationally expensive to compute for every manifold in the list. As a result, we were able to determine these lower bounds only up to $h=6$. 

For the second, computational method, we have used the fact that we must have an identification of line bundles on two diffeomorphic manifolds which leaves certain quantities, related to the line bundle index, invariant. This observation allows searching for suitable $\operatorname{GL}(h,\mathbb{Z})$ transformations, using the algorithm described in 
Section~\ref{sec:search_alg}. This algorithm is applied only on pairs of manifolds that agree at the level of the invariants calculated under method~(i). Fortunately, the invariants used were strong enough to partition the data into small enough classes for $h\leq 6$. The result is an upper bound for the number of diffeomorphism classes for $h\leq 6$. 


The results obtained from applying these methods to the set of simply connected CY three-folds from the KS list are described in Section~\ref{sec:Results} and summarised in Table~\ref{tab:Results} and Figs.~\ref{fig:TriangEquivPlot} and \ref{fig:NumberTriangdivNumberPolytopes}. For $h\leq3$, where computer algebra methods are feasible, the exact number of diffeomorphism classes can be determined. Specifically, for $h=1,2,3$ we find $4$, $27$ and $183$ diffeomorphism classes, respectively. For $h=4,5,6$  we find tight lower and upper bounds for the number of classes, using methods (i) and (ii), with the precise numbers given in Table~\ref{tab:Results}.

An interesting observation from Fig.~\ref{fig:TriangEquivPlot} is that both the (logarithmic) number of FRSTs and diffeomorphism classes depends linearly on $h$. Boldly extrapolating to $h\leq 6$ assuming this dependence, we estimate that a total of $(1.6\pm0.1)\times10^{497}$ FRSTs and a total of between $10^{396}$ and $10^{401}$ diffeomorphism classes can be obtained from the KS list.

There are a number of natural extensions of this work, which we leave for further study. 
One remaining issue is to find an explanation for the $\operatorname{GL}(h,\mathbb{Q})$ transformations discovered between various manifolds. We have no good explanation for the presence of these relations, nor an explanation for why the determinants of these maps are always $\pm1$.

It is clearly desirable to extend our results to higher Picard numbers and to find further invariants which might help to achieve this. On the latter point, we can ask a number of questions. Are there generalisations of the spinor genus to cubic forms? Can one apply spinor genus techniques to the symmetric matrices generated in the course of evaluating invariants in this work? In particular, are there enough, yet unknown, invariants to fully determine the number of diffeomorphism classes? Could these invariants be relevant to questions of boundedness for CY manifolds? It would also be interesting to understand if there is a geometric interpretation for the invariants presented in this work.\footnote{For example, the degree 4 and 6 invariants at the $h=3$ (usually understood in the context of ternary cubic equations, and known as the Aronhold $S$ and $T$ invariants \cite{Banchi_2015}) partially specify the rank and border rank of the associated real ternary cubic. If the degree four invariant vanishes, the polynomial corresponding to the intersection form is a sum of three independent and complex linear forms. If the sextic form is negative, the linear forms are actually real, and if it is positive one is real and two are complex conjugates. In the former case, this means the tensor is `diagonal' over $\mathbb{R}$---in some (potentially non-integral) basis, there are only three intersections $d_{111},d_{222},d_{333}$. It is possible that this has a deeper meaning for our context. Both cases occur in the list, and intriguingly we find integral basis transformations. As an explicit example, we are able to transform the intersection form of the $h=3$ manifold with (polytope \#, triangulation \#) = (1,0) to a diagonal 3-tensor with entries $(-3,-1,1)$.}

\section*{Acknowledgements}
Aditi Chandra was supported by the David Brink fund and Balliol College during the course of this work. Andrei Constantin's research is supported by a Stephen Hawking Fellowship, EPSRC grant EP/T016280/1. Kit Fraser-Taliente is supported by the Gould-Watson Scholarship and Lady Margaret Hall. Thomas Harvey is supported by an STFC studentship. The authors would like to thank Steve Abel, Giulio Gambuti, Naomi Gendler, Liam McAllister, Jakob Moritz, Andreas Schachner and Michael Stillman for useful conversations. We also thank Jonathan Patterson for assistance with running jobs on the ``Hydra" computing cluster belonging to the Oxford Theoretical Physics sub-department.
We thank the authors of Ref.~\cite{CornellCountCY} for useful comments on the first pre-print version of this paper.

\appendix
\section{GCD invariants}
\label{app:basicGCDinvs}

The GCD is an associative and commutative function that ignores signs and zeros:
\be
\begin{aligned}
    \operatorname{GCD}(0,a)&=a~,\\
    \operatorname{GCD}(-a,b) &= \operatorname{GCD}(a,b)~,\\
    \operatorname{GCD}(a,b,c)&=\operatorname{GCD}(\operatorname{GCD}(a,b),c)~,\\
    \operatorname{GCD}(\operatorname{GCD}(a,b),c)&=\operatorname{GCD}(a,\operatorname{GCD}(b,c))~.
\end{aligned}
\ee

{\bfseries Proposition.} The entries of vectors transforming in the fundamental representation of $\operatorname{GL}(h,\mathbb{Z})$ 
have an invariant GCD. Moreover, the GCD provides a complete invariant for the classification of integral vectors (in the fundamental) up to $\operatorname{GL}(h,\mathbb{Z})$ transformations. 
\vspace{4pt}

{\itshape Proof.} Bézout’s identity states that, given a vector $a\in\mathbb{Z}^h$, if $\operatorname{GCD}(\{a_i\}) = d$, then for all $x\in\mathbb{Z}^h$, $a\cdot x = k d$ for integer $k$. Moreover, $k = 1$ is realised for some $x\in\mathbb{Z}^h$. Consider two vectors $v, v'{\in}\mathbb{Z}^h$. Take $\operatorname{GCD}(\{v_i\}){=}d$ and $\operatorname{GCD}(\{v'_j\}){=}d'$. If $v'_j {=} P_j^iv_j$, $\operatorname{GCD}(\{v_j'\}) {=} \operatorname{GCD}(\{P_j^iv_i\}) {=} \operatorname{GCD}(\{k_j d\}){=}d\operatorname{GCD}(\{k_j\})$. This implies that $d$ divides the GCD $d'$ of $v'$. We can apply the same reasoning after swapping the roles of $v$ and~$v'$, instead using the integral matrix $P^{-1}$, and conclude that $d'$ divides $d$. It follows that $d=d'$.

To prove that there is only one invariant of vectors, WLOG, let $\operatorname{gcd}\{v\}=\operatorname{gcd}\left\{v^{\prime}\right\}=1$. Construct $P$ as a matrix with first column $v$, and $Q$ as a matrix with first column~$v^{\prime}$. It is a fact~\cite{niven1991} that these matrices can be chosen to be integral with determinant $\pm1$. Then acting with these matrices on the first unit vector yields $v$ and $v'$ respectively. It follows that $Q P^{-1} v=v^{\prime}$ with $\operatorname{det} Q P^{-1}=1$. 
$\hfill\square$
\vspace{4pt}

An alternative proof of the first part of the above proposition is possible and invites a broader picture which incorporates vectors that do not transform in the fundamental. Once again, let $(v_i)$ be a vector transforming in the fundamental of $\operatorname{GL}(h,\mathbb{Z})$, and $f_v$ be the associated linear form on the dual representation. Consider the set:
\be
S_0 = \left\{f_v(A) \mid A \in H^{1,1}(X,\mathbb{Z})\right\}=\left\{v_in^i \mid n_{i} \in \mathbb{Z}^h\right\}
\ee

Since $\operatorname{GL}(h,\mathbb{Z})$ transformations of determinant $\pm1$ map $\mathbb Z^h$ to itself, it follows that $S_0$ is invariant. If we can construct a well-defined and finite function of this (infinite) set, we can conclude that this function is invariant. The GCD is such a function, due to the following:
\vspace{2pt}

\noindent{\bfseries Property (M).} For $a$ and $b$ integer, and for any integer~$m$, $\operatorname{GCD}(a+mb,b) =\operatorname{GCD}(a,b)$. 
\vspace{2pt}

Using property (M), we can therefore construct an alternative set $\tilde{S}_0$ with the same GCD.
\be
\tilde{S}_0=\left\{v_i \in [1,\dots, h]\right\}~.
\ee

We can follow a similar argument for vectors in other representations. For instance, consider the triple intersection form on $H^{1,1}(X,\mathbb Z)$ and form the sets
\begin{equation}
\begin{aligned}
S_1&=\set{\kappa(A,B,C)\mid A,B,C\in H^{1,1}(X,\mathbb{Z})}~\\
&=\set{d_{ijk}n_1^in_2^jn_3^k\mid n_1,n_2,n_3\in \mathbb{Z}^{h}},\\
S_2&=\set{\kappa(A,A,B)\mid A,B\in H^{1,1}(X,\mathbb{Z})}~\\
&=\set{d_{ijk}n_1^in_1^jn_2^k = d_{i i j}\left(n_1^i\right)^2 n_2^j\\
&~~~~~~~~~~~~~~~~+2 \sum_{i<j} d_{i j k} n_1^i n_1^j n_2^k \mid n_1,n_2\in \mathbb{Z}^{h}},\\
S_3&=\set{\kappa(A,A,A)\mid A\in H^{1,1}(X,\mathbb{Z})}~\\
&=\set{d_{ijk}n_1^in_1^jn_1^k=d_{i i i}\left(n_1^i\right)^3+3 \sum_{i<j}\left[d_{i j j} n_1^i\left(n_1^j\right)^2\right.\\
 &\left.~~~~~~~+d_{i i j}\left(n_1^i\right)^2 n_1^j\right] +6 \sum_{i<j<k} d_{i j k} n_1^i n_1^j n_1^k \mid n_1\in \mathbb{Z}^{h}}~.
\end{aligned}
\end{equation}

Using the GCD properties described above and judicious choices of the $n^i_m$ vectors the sets get reduced to:
\be
\begin{aligned}
\tilde{S}_1 &=\left\{d_{i j k} \mid i, j, k =1, \ldots, h\right\}, \\
\tilde{S}_2 &=\left\{d_{i i j} \mid i, j=1, \ldots, h\right\}\cup\left\{2 d_{i j k} \mid i, j, k=1, \ldots, h\right\}, \\
\tilde{S}_3 &=\left\{d_{i i i} \mid i{=}1, \ldots, h\right\} \cup\left\{3\left(d_{i i j} \pm d_{i j j}\right) \mid i, j {=}1, \ldots, h\right\} \\
& \quad\cup\left\{6 d_{i j k} \mid i, j, k=1, \ldots, h\right\}.
\end{aligned}
\ee

These new finite sets have the same GCDs as $S_1$, $S_2$ and $S_3$. We also note that it is possible to apply the same approach to four-arrays \cite{4FoldsGCDsJamesGray}, such as $b_{ijkl}=c_{(i}d_{jkl)}$, by starting with infinite sets in the same way as above, and reducing them to finite ones with the same GCD. In this way, the following sets are obtained, whose GCD correspond to H\"ubsch invariants~\cite{hubschCYB}:
\be
\begin{aligned}
\tilde{S}_4 & =\left\{b_{i j k l} \mid i, j, k, l=1, \ldots, h\right\} \\
\tilde{S}_5 & =\left\{b_{i i j k} \mid i, j, k=1, \ldots, h\right\} \\
&\quad\cup\left\{2 b_{i j k l} \mid i, j, k, l=1, \ldots, h\right\}, \\
\tilde{S}_6 & =\left\{b_{i i i j} \mid i, j=1, \ldots, h\right\}\\
&\quad\cup\left\{3\left(b_{i i j k} \pm b_{i j j k}\right) \mid i, j, k=1, \ldots, h\right\} \\
& \quad\cup\left\{6 b_{i j k l} \mid i, j, k, l=1, \ldots, h\right\} \\
\tilde{S}_7 & =\left\{b_{i i i i} \mid i=1, \ldots, h\right\} \\
&\quad\cup\left\{6 b_{i i j j} \pm 4\left(b_{i i i j} \pm b_{i j j j}\right) \mid i, j=1, \ldots, h\right\} \\
& \quad\cup\left\{12\left(b_{i j k k} \pm b_{i j j k} \pm b_{i i j k}\right) \mid i, j, k=1, \ldots, h\right\}\\
& \quad\cup\left\{24 b_{i j k l} \mid i, j, k, l=1, \ldots, h\right\} .
\end{aligned}
\ee
Note that in the main text, we used the notation $\tilde S_i$ for the GCDs of the above sets. 

It is also possible to apply the same approach to arrays with reduced symmetry, in which case the invariants become significantly more complicated to write down. Finally, for totally symmetric arrays we can also consider the coefficients in the polynomial $d_{ijk}t^it^jt^k$, which differ from $d_{ijk}$ simply by the relevant binomial coefficient, which is its multiplicity in the array. These coefficients also transform in the same representation, but with different weights corresponding to the binomial coefficients. Crucially, the transformation matrix remains integral and of determinant $\pm1$. This `binomial-adjusted GCD' is very similar, but not identical, to $\tilde{S}_3$. The same invariant can also be written for the 4-array $b_{ijkl}$, and the result looks similar to $\tilde{S}_7$:

\be\begin{aligned}
\tilde{S}_3' &=\left\{d_{i i i} \mid i=1, \ldots, h\right\} \cup\left\{3\left(d_{i i j}\right) \mid i, j =1, \ldots, h\right\} \\
& \quad\cup\left\{6 d_{i j k} \mid i, j, k=1, \ldots, h\right\}~,\\
\tilde{S}_7' & =\left\{b_{i i i i} \mid i=1, \ldots, h\right\} \cup\left\{6 b_{i i j j} \mid i, j=1, \ldots, h\right\} \\
&\quad\cup\left\{4 b_{i i i j} \mid i,j{=}1, \ldots, h\right\} \cup\left\{12b_{i j k k}\mid i, j, k{=}1, \ldots, h\right\}\\
& \quad\cup\left\{24 b_{i j k l} \mid i, j, k, l=1, \ldots, h\right\}.
\end{aligned}
\ee

Of course, the same procedure can be applied for higher tensor powers of $\mathbb{R}$ and $\mathbb{H}$.

\subsubsection*{Polynomial GCD invariants}

We examine the invariant sets of the polynomial representation quadratic in the entries of the intersection form. Consider the new set:
\be Q_0=\left\{\kappa(A,B,C)\kappa(D,E,F)\mid A,\dots,F\in H^{1,1}(X,\mathbb{Z})\right\}\ee

Again, such a set will have an invariant GCD. Unfortunately, all GCDs will factorise, and no new information arises from the set of quadratic monomials. However, the following sets can, in principle (and in practice, do) give new information:
\be
\begin{aligned}
Q_1&=\left\{\kappa\left(A,B,C\right)\kappa\left(D,E,F\right)-\kappa\left(A,B,D\right)\kappa\left(C,E,F\right)\mid \right.\\
&\qquad\left. A,\dots,F\in H^{1,1}(X,\mathbb{Z})\right\}~,\\
Q_2&=\left\{\kappa\left(A,B,C\right)\kappa\left(D,E,F\right)+\kappa\left(A,B,D\right)\kappa\left(C,E,F\right)\mid\right.\\ 
&\left.\qquad A,\dots,F\in H^{1,1}(X,\mathbb{Z})\right\}~.
\end{aligned}
\ee

In Picard number 2, the quadratic representation splits into a 3-dimensional and a 7-dimensional representation, which correspond to the $\operatorname{Sym}^2\mathbf{R}$ and $\wedge^2\mathbf{R}$ representations. That is, the first infinite set we constructed actually has two different infinite subsets, each of which is invariant under the basis transformation and therefore can be acted on with a function like the GCD. As above, these sets can be turned into finite sets.

Similarly, the set $\set{\kappa(A,B,C)\kappa(D,E,F)\kappa(H,I,J)}$ will break into several infinite subsets, with possibly-unrelated GCDs. These sets can be identified using the representation algebra (e.g. using $\mathtt{LiE}$), and constructed explicitly using an appropriate (anti-)symmetrisation procedure. Now we can also consider the analogous versions of $\tilde{S}_2$ and $\tilde{S}_3$, by taking sets of the form 
\be 
\label{eq:polyeq}
\begin{aligned}
Q_1'&=\left\{\kappa\left(A,B,C\right)\kappa\left(D,D,E\right)-\kappa\left(A,B,D\right)\kappa\left(C,D,E\right)\mid\right.\\ 
&\left.\qquad A,\dots,F\in H^{1,1}(X,\mathbb{Z})\right\}
\end{aligned}
\ee
and then trimming them down to a finite set with the same GCD.

\subsubsection*{A useful polynomial representation}

As discussed above, it is possible to construct polynomial representations explicitly using an appropriate (anti-)symmetrisation procedure. Unfortunately, this method of constructing representations---whilst complete---rapidly becomes unworkable, as the lowest-degree representations are rarely low-dimensional, as they are polynomial in the dimension of $\mathbf{R}$, and do not seem to extract much new information. One representation which seems to be more useful is the degree-$h$ $h$-array from Eq.~\eqref{eq:SymDet}, which we term the Hessian-determinant representation. It has an invariant GCD (as well as a binomially-weighted GCD):
\be
\label{eq:hessGCD}
\operatorname{GCD}(\{A_{i_1\dots i_h}\})~.
\ee

This is a degree-$h$ polynomial in $h$ variables $z^i$, which corresponds to a $\binom{2h-1}{h}$-dimensional representation. The coefficients of each monomial (which are functions of the $d_{ijk}$ integers) form the entries for the representation. For $h=2$, this is the three-dimensional quadratic polynomial representation; for $h=3$, it is already ten-dimensional. For $h=2$, the Hessian representation coincides with the antisymmetrised quadratic representation, and has basis vectors
\be
ad-bc,\; b^2-ac,\; c^2-bd~.
\ee

Surprisingly, this Hessian GCD invariant provides very reasonable discrimination in cases where we do not wish to evaluate higher degree invariants. Note that it also has descendent GCDs analogous to $\tilde{S}_2$ and $\tilde{S}_3$, and a binomially-weighted GCD analogous to $\tilde{S}_3'$.

\section{Hyperdeterminants and polynomial invariants}
\label{app:polyinvs}

One generalisation of the Cayley hyperdeterminant, the Gelfand hyperdeterminant \cite{GelfandDiscrim}, is actually significantly more powerful than strictly necessary in our consideration. They---and potentially other elements of the invariant ring---are invariant under simultaneous independent changes of basis on the left and on the right. As mentioned in the main text, various different generalisations of the Cayley hyperdeterminant are possible which suggest various strategies for constructing invariants, but at least three of these coincide for $2\times 2\times 2$ arrays\footnote{The three approaches we termed the Gelfand hyperdeterminant, the Pascal hyperdeterminant and the `SymDet'.}. This must happen, as there is only one polynomial invariant. The exact form of the Gelfand hyperdeterminant is only known for two-dimensional 3-tensors (and partially known for three-dimensional 3-tensors \cite{sl3cfundamentalinvariants}). However, it is known to exist for any $h\times h \times h$ array with degree :
\be\operatorname{deg}(h)=\sum_{0 \leq j \leq (h+1)/ 2} \frac{(j+(h+1)+1) !}{j !^3((h+1)-2 j) !} \cdot 2^{(h+1)-2 j}\ee

The degree of this polynomial grows exponentially with~$h$, taking the following values up to Picard number five: $1,4,36,272,$ and $2070$. Anything beyond $h=3$ is unworkable.

\subsubsection*{Syzygies}
The existence of syzygies in the cases we are interested in can be determined by counting the number of singlets in the group-theoretic decomposition up to some degree. If this number (removing simple monomial powers of invariants) exceeds the number of basis-independent degrees of freedom, it will be clear there must be a syzygy. No syzygies were encountered in the course of calculating the invariants in this paper, but they are known to exist in similar cases \cite{GelfandDiscrim}.

\subsubsection*{Combinatorics for graphs}
When is it possible to write down a polynomial invariant that does not vanish? Consider $\delta$ copies of the intersection numbers $d_{i_1,j_1,k_1}, \dots, d_{i_\delta,j_\delta,k_\delta}$. We need to contract the $3\delta$ indices in such a way that the result does not vanish, which means we need $3\delta/h$ copies of the Levi-Civita symbol in $h$-dimensions. Then degree-$\delta$ invariants of this form will only exist when $3\delta/h \in \mathbb{Z}$. No two of $i_1,j_1,k_1$ should ever be contracted with the same $\epsilon$-tensor, as this would cause the invariant to vanish. Similarly, no two tensors $d_{i_1,j_1,k_1}$ should ever have exactly the same $\epsilon$-contractions. This means we need at least $\delta$ different ways of selecting 3 Levi-Civita symbols from our list of $3\delta/h$, giving a bound: $\delta \leq 3\delta/h$. Interestingly, this bound is saturated by $h$ the $k$th triangle number and $\delta= \binom{k+2}{3}$. The four lowest such incidences are $(h,\delta)=$ (1, 1), (3, 4), (6, 10), (10, 20). These are precisely the `coincidental' invariants appearing in Table~\ref{tabdegnum}, and correspond to the case when there is exactly one admissible (and possibly nonvanishing) contraction.

We can think of these contraction patterns as bipartite graphs with edges running from the $3\delta/h$, $h$-valent $\epsilon$-type vertices to the $\delta$ trivalent array-type nodes. Due to the symmetries of the 3-tensor and Levi-Civita symbol, the topology of this graph uniquely specifies (up to sign) a contraction pattern and hence an invariant. It is easy to exhaustively construct all such graphs for a given $\delta$, $h$, and it is particularly easy when there is just one possible contraction pattern (when the bound is saturated). If we attempt to detect the Pascal hyperdeterminant for $h = 3$, we generate $330$ different contraction patterns, of which just four are topologically distinct with approximately equal incidences. Two of those four yield the Pascal hyperdeterminant, and the others yield an invariant identically equal to zero. For $h=2$, $\delta=4$, there are two graphs, leading to one zero invariant and one non-zero. One can then directly evaluate the invariants, which is still practicable up to $h=6$. All invariants of low degree mentioned above are constructible with this method, but if the degree gets too high, it becomes unworkable. Nevertheless, sparse array methods in $\mathtt{Mathematica}$, coupled with optimised index contraction paths, ensure that these invariants remain calculable for the Picard numbers considered in this paper. 

\begin{figure}
    \centering
    \includegraphics[width=0.9\linewidth]{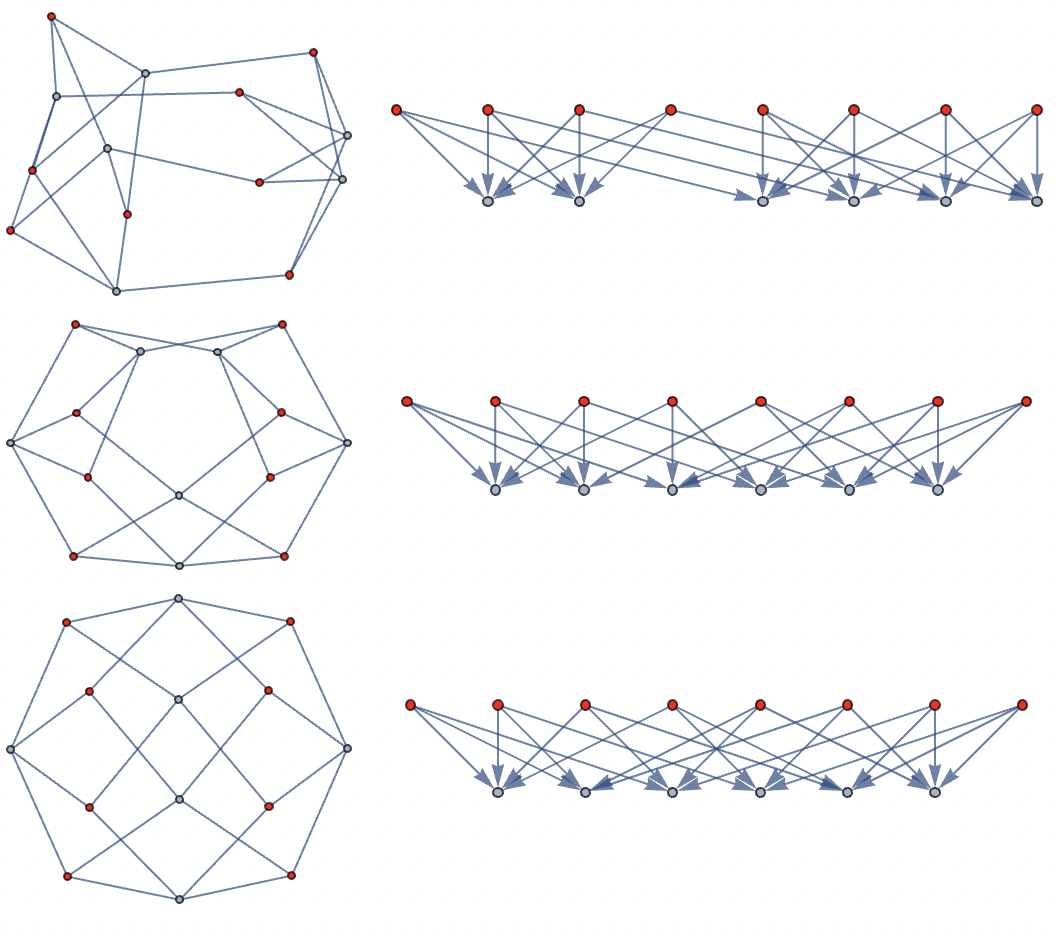}
    \caption{Contraction patterns for degree 6, $h=3$ invariants which evaluate to the non-zero Pascal/$2h$-invariant. Each graph is shown once in directed and once in undirected form. Levi-Civita-type vertices are coloured red.}
    \label{fig:non-zero3inv}
\end{figure}

\begin{figure}
    \centering
    \includegraphics[width=0.9\linewidth]{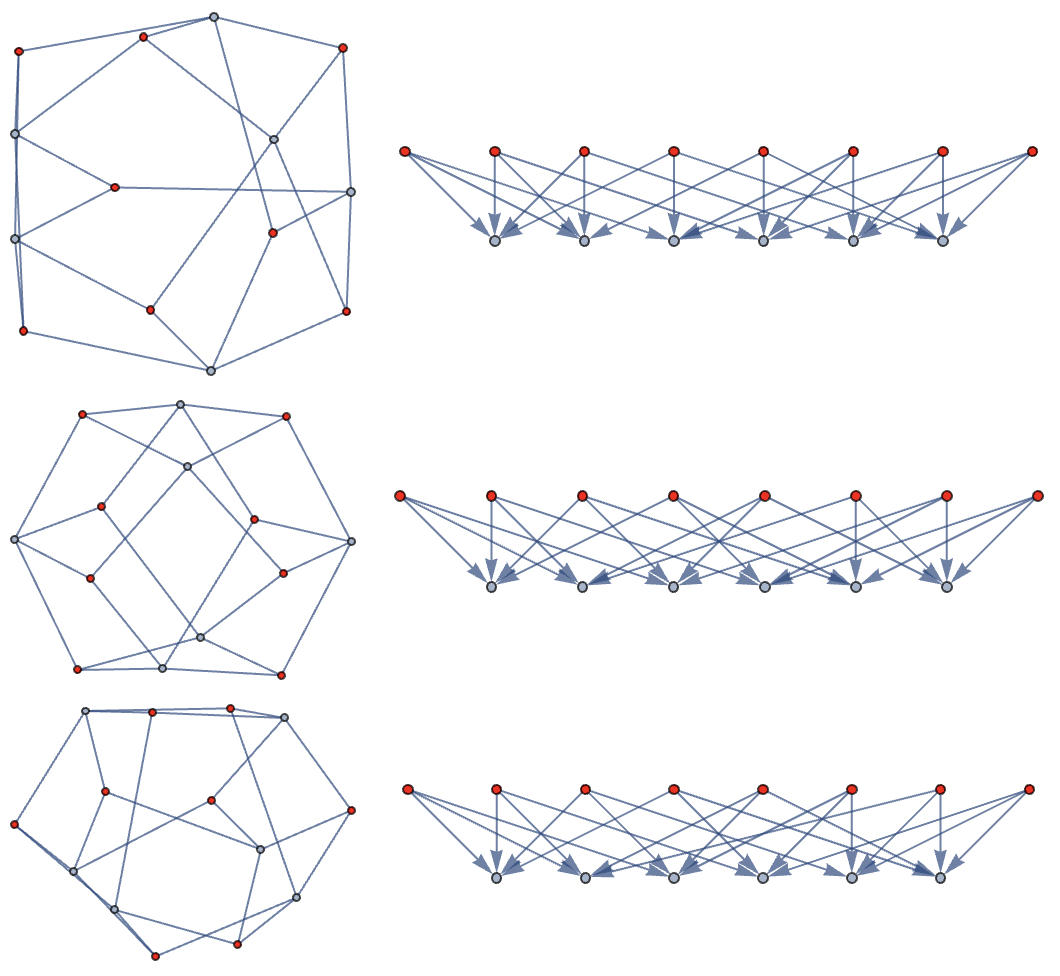}
    \caption{Contraction patterns for degree 6, $h=3$ invariants which evaluate to zero. Each graph is shown once in directed and once in undirected form. Levi-Civita-type vertices are coloured red.}
    \label{fig:zero3inv}
\end{figure}

We include for interest the 6 different graph topologies for $h=3$. Of the corresponding invariants, three (shown in Fig.~\ref{fig:non-zero3inv}) evaluate to zero and three to non-zero (shown in Fig.~\ref{fig:zero3inv}). As before, they are presented in directed and undirected form, and $\epsilon$-type vertices are coloured red. For higher tensors (e.g. 4-tensors), or higher degrees/$h$, the initial construction of all possible graphs becomes hard. However, as nonzero contraction patterns are not `rare' in the space of admissible bipartite graphs, it is often sufficient to sample the space of bipartite graphs of fixed valence, rejecting graphs that vanish by symmetry.

\subsubsection*{Additional polynomial invariants generated from the Hessian-determinant representation}

From the Hessian-determinant representation we can construct the four-array $C$ with indices $I$ valued in~$\operatorname{Sym}^{n/4}\mathbf{H}$:
\be
C_{IJKL} = A_{(i_1\dots i_{n/4})(i_{n/4+1}\dots i_{n/2})(i_{n/2+1}\dots i_{3n/4})(i_{3n/4+1}\dots i_{n})}
\ee
Then take the PDET of $C_{IJKL}$. We did not evaluate this invariant as, in doing so, we regain the computational complexity issues associated to taking the PDET for large dimension. Similarly, for $h$ divisible by 3 we could symmetrise appropriately and then take the PDET of a new 3-tensor.

\subsubsection*{Additional polynomial invariants for small Picard number}
Finally, we consider some other invariants which can be constructed but vanish for  $h$ larger than 3, 4, or 5 due to a non-maximal generic rank or generalised tensor rank. The first two invariants are algebraically related to the other known polynomial invariants. We include these to indicate that there are some constructions which give invariants but become zero after some point.
\begin{enumerate} 
\item An invariant which vanishes for $h\geq3$: 
\be
\label{eq:d2syminv}
\begin{aligned}
A_{ijklmn}&=A_{(ijklmn)}=d_{(ijk}d_{lmn)}\\
B_{IJ} &= A_{(ijk)(lmn)} \rightarrow \det{B}~.
\end{aligned}
\ee
This coincides with the hyperdeterminant$^2$ for its one nontrivial case. One can also calculate the rank and signature of $B$.

\item An invariant which vanishes for $h\geq 4$: 
\be
\label{eq:d4syminv}
\begin{aligned}
    A'_{i_1\dots i_{12}}&=A'_{(i_1\dots i_{12})}=d_{(i_1i_2i_3}d_{i_4i_5i_6}d_{i_7i_8i_9}d_{i_{10}i_{11}i_{12})} \\
B'_{IJ} &= A'_{(i_1\dots i_{6})(i_7\dots i_{12})} \rightarrow \det{B'}~.
\end{aligned}
\ee
This coincides with hyperdeterminant$^7$ for $h=2$, and is polynomial in $I_{4,3}$ and $I_{6,3}$ for $h=3$. One can also calculate the rank and signature of $B'$.
\item An invariant which vanishes for $h\geq 5$: 
\be
\label{eq:pdetdc}
\operatorname{PDET}(d_{(ijk}c_{2,l)})~.\ee

Without symmetrising, it vanishes. If, instead, one antisymmetrises and then evaluates the PDET, one has $\operatorname{PDET}(d_{ij[k}c_{2,l]})$ which vanishes for $h\geq3$.
\end{enumerate}

\section{Generating polynomial invariants}
\label{app:GeneratingPolWithLA}
It is interesting to note that $\operatorname{GL}(h,\mathbb{Z})$ is finitely generated \cite{deLaHarpeTopicsGeoGroup} by two (or three) matrices $S,T,U$ (depending on dimension). $S$ and $T$ suffice for $h$ even, and all three are required for $h$ odd\footnote{Note that we take the approach that given a generator, we are also permitted to use its inverse.}.

\begin{small}
\begin{align*}
S&= \left(\begin{array}{cccccc}
1 & 1 & 0 & \ldots & 0 & 0 \\
0 & 1 & 0 & \ldots & 0 & 0 \\
0 & 0 & 1 & \ldots & 0 & 0 \\
\ldots & \ldots & \ldots & \ldots & \ldots & \ldots \\
0 & 0 & 0 & \ldots & 1 & 0 \\
0 & 0 & 0 & \ldots & 0 & 1
\end{array}\right)~,  \\[4pt]
T &=\left(\begin{array}{cccccc}
0 & 0 & 0 & \ldots & 0 & 1 \\
1 & 0 & 0 & \ldots & 0 & 0 \\
0 & 1 & 0 & \ldots & 0 & 0 \\
\ldots & \ldots & \ldots & \ldots & \ldots & \ldots \\
0 & 0 & 0 & \ldots & 0 & 0 \\
0 & 0 & 0 & \ldots & 1 & 0
\end{array}\right)~, \\[4pt]
U&=\left(\begin{array}{cccccc}
-1 & 0 & 0 & \ldots & 0 & 0 \\
0 & 1 & 0 & \ldots & 0 & 0 \\
0 & 0 & 1 & \ldots & 0 & 0 \\
\ldots & \ldots & \ldots & \ldots & \ldots & \ldots \\
0 & 0 & 0 & \ldots & 1 & 0 \\
0 & 0 & 0 & \ldots & 0 & 1
\end{array}\right)~.
\end{align*}
\end{small}

If we find the matrix representations $R^d_S, R^d_T, R^d_U$ of the generators in the relevant degree-$\delta$ polynomial representation, the non-zero simultaneous eigenvectors correspond to singlets. This immediately enables us to find the lowest order invariants for small $h$. Unfortunately, these matrices become exponentially large.

Firstly, one can consider the `torus invariants'. Consider the subgroup $D$ of $\operatorname{SL}(h,\mathbb{R})$ given by the invertible diagonal matrices. The action of $D$, known as the maximal torus, maps monomials into monomials. We know that under $P\in \operatorname{GL}(h,\mathbb{R})$ a degree-$\delta$ invariant $I_{\delta,h}$ of the 3-tensor, must be mapped to $\det{P}^{3\delta/h} I_{\delta,h}$. It follows that the monomials present in the invariant must be those which transform with a factor of $\prod_{i=1}^h t_i^g$ under $D$, for $g = 3\delta/h$. This heavily restricts the number of admissible monomials, and when considered using the Lie algebra formalism these `torus monomials' are precisely those with weight zero. Moreover, each monomial should come with its entire permutation orbit, and thus we can combine multiple torus monomials.

Consideration of the torus-invariant monomials also lets us determine which degree plethysms could actually entertain invariants. If invariants transform as $(\det{P})^{3\delta/h}$, as they are polynomial representations, we must have $3\delta/h$ an integer, $k$. Then the only admissible degrees are those $\delta$ given by $kh/3$ for integer $k$, a condition which is naturally very closely related to the Levi-Civita graph discussion above. As determined above, not all of these admissible degrees actually support invariants. The eigenvector operation $\texttt{eigenvectors}(R_S)$ can be converted to a $\texttt{nullSpace}$ operation, and then we can therefore project one side of the matrix ($R_S-\operatorname{Id}$) onto the admissible weight zero monomials. It turns out to be sufficient to consider only the eigenvectors of the $S$ matrix (we can neglect $T$ and $U$). For actual evaluation of polynomial invariants, it is much faster to directly evaluate them using the PDET/Levi-Civita formulae, rather than substituting into a symbolic expression. 

\section{Data formatting}
\label{app:data}
\begin{table}
  \scriptsize
  \centering
  \begin{tabular}{|l|l|l|}
    \hline
    \textbf{Name} & \textbf{Explanation} & $h$ \\
    \hline
    $\texttt{"2h-invariant-X-ic"}$  & Eq.~\ref{eq:PDET} & 2-5\\ \hline
    $\texttt{"GCD-dijk"}$  & $\tilde{S}_1$ in Eq.~\ref{eq:Hubsch:St0to3} & 2-6 \\ \hline
    $\texttt{"GCD-diij-St\_2"}$ & $\tilde{S}_2$ in Eq.~\ref{eq:Hubsch:St0to3}  & 2-6 \\ \hline
    $\texttt{"GCD-diij-St\_3"}$ & $\tilde{S}_3$ in Eq.~\ref{eq:Hubsch:St0to3} & 3-6 \\ \hline
    $\texttt{"hessGCD"}$  & Eq.~\ref{eq:hessGCD} & 2-6 \\ \hline
    $\texttt{"SVtoVrank"}$  & $\rk$ of $d$ as $\operatorname{Sym}^2\mathbf{H}\rightarrow \mathbf{H}$ & 2-6 \\ \hline
    $\texttt{"GCD Chern"}$  & $\tilde{S}_0$ in Eq.~\ref{eq:Hubsch:St0to3} & 2-6 \\ \hline
    $\texttt{"d\^{}4DetandRankSig"}$  & \{det, rk, sig\} of Eq.~\ref{eq:d4syminv} & 2\\ \hline
    $\texttt{"d\^{}2DetandRankSig"}$  & Eq.~\ref{eq:d2syminv} & 2\\ \hline
    $\texttt{"c2dinvariantsSymandAntisym"}$ & Eq.~\ref{eq:pdetdc}, \{sym, antisym\} & 2\\ \hline
    $\texttt{"GCDofCDwithBinom"}$  & $\{\tilde{S}_4,\;\tilde{S}_7'\}$ of Eqs.~\ref{eq:Hubsch:St4to7}, \ref{eq:HubschStp3Stp7} & 2-6 \\ \hline
    $\texttt{"GCDofCDASym"}$ & $\tilde{S}_4$' of Eq.~\ref{eq:HubschSt4p} & 2-6 \\ \hline
    $\texttt{"GCDofCD2same"}$  & $\tilde{S}_5$ of Eq.~\ref{eq:Hubsch:St4to7} & 2-6 \\ \hline
    $\texttt{"GCDofCD3same"}$  & $\tilde{S}_6$ of Eq.~\ref{eq:Hubsch:St4to7} & 2-6 \\ \hline
    $\texttt{"GCDofCD4same"}$  & $\tilde{S}_7$ of Eq.~\ref{eq:Hubsch:St4to7} & 2-6 \\ \hline
    $\texttt{"BinomialDijkGCD"}$  & $\tilde{S}_7'$ of Eq.~\ref{eq:HubschStp3Stp7} & 2-6 \\ \hline
    $\texttt{"GCDCubicTotalSym"}$  & GCD of $S^3\mathbf{R}$ & 2-4 \\ \hline
    $\texttt{"GCDCubic6D"}$  & GCD of 4D rep in $\mathbf{R}^3$ & 2 \\ \hline
    $\texttt{"GCDCubic4D"}$  & GCD of 6D rep in $\mathbf{R}^3$ & 2\\ \hline
    $\texttt{"detS2VtoS2VRankSignature"}$  & $\det$, $\rk$, sig of ${\hat{b}}$ from \ref{subseq:detrk} & 3 \\ \hline
    $\texttt{"GCDQuadraticSym"}$  & GCD of $S^2\mathbf{R}$ & 2-4 \\ \hline
    $\texttt{"GCDQuadraticAsym"}$  & GCD of $\wedge^2\mathbf{R}$ & 2-4 \\ \hline
    $\texttt{"GCDCubic10D"}$  & GCD of 10D rep in $\mathbf{R}^3$ & 3 \\ \hline
    $\texttt{"PDETCxDSym"}$ & Eq.~\ref{eq:pdetdc}, symmetric only & 3, 4 \\ \hline
    $\texttt{"CEpsDeg8Invariant"}$ & $I_{(8,6),5}(d_{ijk},c_{2,l})$ in \ref{sec:graphs} & 5 \\ \hline
    $\texttt{"RankAndSignatureS2V->S2V"}$ & $\rk$ and $\operatorname{sig}\hat{b}$ from \ref{subseq:detrk} & 5,6 \\ \hline
    $\texttt{"SymDetDeg\#\#andRankSig"}$ & \{det, rk, sig\} of Eq.~\ref{eq:SymDet} & 4,6 \\ \hline
    $\texttt{"CEpsDeg10Invariant"}$ & $I_{(10,6),6}(d_{ijk},c_{2,l})$ in \ref{sec:graphs} & 6 \\ \hline
    $\texttt{"I(4,4),4MixedInvariant"}$ & $I_{(4,4),4}(d_{ijk},c_{2,l})$ in \ref{sec:graphs} & 4 \\ \hline
  \end{tabular}
  \caption{Table giving the names given to each invariant in the data, an explanation of what the name refers to, and for which Picard numbers the invariant was evaluated. We note that not all obvious or accessible invariants were evaluated, particularly for $h=2,3$ where we had perfect discrimination.}
  \label{tab:mergedInvars}
\end{table}
For each Picard number $h\leq 6$, there are two files in the data, found on the project site\footnote{\texttt{http://www-thphys.physics.ox.ac.uk/}

\hspace{0.35cm}\texttt{projects/CalabiYau/KSEquiv}}. One is the list of FRSTs (contained in $\texttt{ManifoldData.zip}$), with their associated data. This is formatted as a JSON file in the following format:
\begin{equation*}
\begin{aligned}
&\texttt{\{\{"manifoldIndex": \#,"polytope": \#,}\\
&\texttt{"triangulation": \#,"intersec": \#,"c2": \#,}\\
&\texttt{"invariant1": \#,"invariant2": \#, \dots,}\\
&\texttt{"SimplyConnected": \#\}},
\\&\texttt{\dots \}}
\end{aligned}
\end{equation*}
Here, $\texttt{manifoldIndex}$ is just an index enumerating the position of a FRST in the list of favourable FRSTs generated by $\texttt{cytools}$. Moreover, $\texttt{cytools}$ yields the $\texttt{polytope}$ and $\texttt{triangulation}$ numbers. For ease of use, all FRSTs are referred to by a triple of indices: $\texttt{\{manifoldIndex, polytope, triangulation\}}$ = $\texttt{\{m, p, t\}}$. Any invariants appearing in these files (for each $h$) were used in the determination of the lower bounds in Table~\ref{tab:Results}. We list which invariants were calculated in Table~\ref{tab:mergedInvars}. In this work, we only considered manifolds with $\texttt{SimplyConnected}=$True.

The next file (contained in $\texttt{EquivalenceData.zip}$) is composed of three lists and gives the equivalence data. They are formatted as $\texttt{Mathematica}$ lists. The first is a list of classes of FRSTs which all have the same invariants (for the cases $h=2,3$, these classes are all singlets, as $\operatorname{GL}(h,\mathbb{Z})$ relations have been explicitly ruled out). Each FRST is listed using our three-index $\texttt{\{m, p, t\}}$. The next is an association of linear maps between FRSTs, taking pairs of FRSTs to lists of transformation matrices (along with their determinants). The last is an association describing the FRSTs with numerically duplicate data, taking single 3-indices $\texttt{\{m,p,t\}}$ to lists of 3-indices which also have the same data. To illustrate this, describe a simple dataset of five manifolds, where we have searched for transformation matrices with entries $\leq 3$. We keep the labelling indices $\texttt{\{m,p,t\}}$ generic.

\texttt{\{"SameInvClasses/UniqueManifoldsWithEntriesin3" -> \{\{\{m1, p1, t1\},\{m5, p5, t5\}\}, \{\{m3, p3, t3\}\}\}, \\
 "Assoc of transformation matrices" -> <|\{\{m1, p1, t1\}, \{m2, p2, t2\}\}-> \{\{$P_1$, $\det P_1 $\},\{$P_2$, $\det P_2$\}\}|>, \\
 "Duplicate PolyTriangs Assoc" -> <| \{m3, p3, t3\} -> \{\{m3, p3, t3\}, \{m4, p4, t4\}\}|>\}}
 
We refer to the manifold $\texttt{\{mi, pi, ti\}}$ as $m_i$. We see that $m_3$ has data duplicate to $m_4$. Having searched for transformation matrices with entries $\leq 3$, we have identified that the data in $m_1$ can be related to $m_2$ by multiplication by either the matrices $P_1$ or $P_2$. However, we have found no relation between $m_1$ and $m_5$. $m_3$ has manifestly different invariants and is therefore in a different list to $m_1$ and $m_5$. Consequently, as they have the same set of invariants, both remain in the list $\texttt{"SameInvClasses/UniqueManifoldsWithEntriesin3"}$.

\bibliography{bibliography}

\providecommand{\href}[2]{#2}\begingroup\raggedright\begin{thebibliography}{10}

\bibitem{WALL1966}
C.~Wall, ``Classification problems in differential topology. v. on certain
  6-manifolds.'' Inventiones mathematicae {\bfseries 1} (1966) 355--374.
  \url{http://eudml.org/doc/141839}.

\bibitem{Tian:1987}
G.~Tian, ``{Smoothness of the Universal Deformation Space of Compact Calabi-Yau
  Manifolds and its Petersson-Weil Metric},'' in {\em {Mathematical Aspects of
  String Theory}}, {Yau, S.-T.}, ed., pp.~629--646.
\newblock World Scientific, 1987.

\bibitem{TodorovL1989}
A.~Todorov, ``{The Weil-Petersson Geometry of the Moduli Space of $SU(n \geq
  3)$ (Calabi-Yau) Manifolds I},'' Commun. Math. Phys. {\bfseries 126} (1989)
  325--346.

\bibitem{Gross:1997}
M.~Gross, ``{The deformation space of Calabi-Yau $n$-folds can be
  obstructed},'' in {\em {Mirror symmetry II}}, pp.~401--411.
\newblock Amer. Math. Soc,, 1997.
\newblock \href{http://arxiv.org/abs/9402014}{[arXiv:9402014 [alg-geom]]}.

\bibitem{Grassi1991}
A.~Grassi, ``{On minimal models of elliptic threefolds},'' Mathematische
  Annalen {\bfseries 290} (1991) 287--301.

\bibitem{Gross1994}
M.~Gross, ``{A finiteness theorem for elliptic Calabi-Yau threefolds},'' Duke
  Mathematical Journal {\bfseries 74} (1994) 271--299.

\bibitem{Wilson2017BoundednessCYs}
P.~M.~H. {Wilson}, ``{Boundedness questions for Calabi-Yau threefolds},''
  \href{http://dx.doi.org/10.48550/arXiv.1706.01268}{arXiv e-prints (June,
  2017) arXiv:1706.01268},
  \href{http://arxiv.org/abs/1706.01268}{[arXiv:1706.01268 [math.AG]]}.

\bibitem{Vafa:2005ui}
C.~Vafa, ``{The String landscape and the swampland},''
  \href{http://arxiv.org/abs/hep-th/0509212}{[arXiv:hep-th/0509212]}.

\bibitem{Palti:2019pca}
E.~Palti, ``{The Swampland: Introduction and Review},''
  \href{http://dx.doi.org/10.1002/prop.201900037}{Fortsch. Phys. {\bfseries 67}
  no.~6, (2019) 1900037},
  \href{http://arxiv.org/abs/1903.06239}{[arXiv:1903.06239 [hep-th]]}.

\bibitem{vanBeest:2021lhn}
M.~van Beest, J.~Calder\'on-Infante, D.~Mirfendereski, and I.~Valenzuela,
  ``{Lectures on the Swampland Program in String Compactifications},''
  \href{http://dx.doi.org/10.1016/j.physrep.2022.09.002}{Phys. Rept. {\bfseries
  989} (2022) 1--50}, \href{http://arxiv.org/abs/2102.01111}{[arXiv:2102.01111
  [hep-th]]}.

\bibitem{Batyrev:1993dm}
V.~V. Batyrev, ``{Dual Polyhedra and Mirror Symmetry for Calabi-Yau
  Hypersurfaces in Toric Varieties},'' J.Alg.Geom. {\bfseries 3} (1994) ,
\href{http://arxiv.org/abs/alg-geom/9310003}{[arXiv:alg-geom/9310003
  [alg-geom]]}.

\bibitem{kreuzer2000complete}
M.~Kreuzer and H.~Skarke, ``{Complete classification of reflexive polyhedra in
  four-dimensions},'' \href{http://dx.doi.org/10.4310/ATMP.2000.v4.n6.a2}{Adv.
  Theor. Math. Phys. {\bfseries 4} (2000) 1209--1230},
  \href{http://arxiv.org/abs/hep-th/0002240}{[arXiv:hep-th/0002240]}.

\bibitem{Candelas:2012uu}
P.~Candelas, A.~Constantin, and H.~Skarke, ``{An Abundance of K3 Fibrations
  from Polyhedra with Interchangeable Parts},''
  \href{http://dx.doi.org/10.1007/s00220-013-1802-2}{Commun. Math. Phys.
  {\bfseries 324} (2013) 937--959},
\href{http://arxiv.org/abs/1207.4792}{[arXiv:1207.4792 [hep-th]]}.

\bibitem{Candelas:2016fdy}
P.~Candelas, A.~Constantin, and C.~Mishra, ``{Calabi-Yau Threefolds with Small
  Hodge Numbers},'' \href{http://dx.doi.org/10.1002/prop.201800029}{Fortsch.
  Phys. {\bfseries 66} no.~6, (2018) 1800029},
  \href{http://arxiv.org/abs/1602.06303}{[arXiv:1602.06303 [hep-th]]}.

\bibitem{Batyrev:2005jc}
V.~Batyrev and M.~Kreuzer, ``{Integral cohomology and mirror symmetry for
  Calabi-Yau 3-folds},''
  \href{http://arxiv.org/abs/math/0505432}{[arXiv:math/0505432]}.

\bibitem{Demirtas:2020dbm}
M.~Demirtas, L.~McAllister, and A.~Rios-Tascon, ``{Bounding the Kreuzer-Skarke
  Landscape},'' \href{http://dx.doi.org/10.1002/prop.202000086}{Fortsch. Phys.
  {\bfseries 68} (2020) 2000086},
  \href{http://arxiv.org/abs/2008.01730}{[arXiv:2008.01730 [hep-th]]}.

\bibitem{Demirtas:2018akl}
M.~Demirtas, C.~Long, L.~McAllister, and M.~Stillman, ``{The Kreuzer-Skarke
  Axiverse},'' \href{http://dx.doi.org/10.1007/JHEP04(2020)138}{JHEP {\bfseries
  04} (2020) 138}, \href{http://arxiv.org/abs/1808.01282}{[arXiv:1808.01282
  [hep-th]]}.

\bibitem{BrodieGeodesicExtKahler}
C.~R. Brodie, A.~Constantin, A.~Lukas, and F.~Ruehle, ``{Geodesics in the
  extended K\"ahler cone of Calabi-Yau threefolds},''
  \href{http://dx.doi.org/10.1007/JHEP03(2022)024}{JHEP {\bfseries 03} (2022)
  024}, \href{http://arxiv.org/abs/2108.10323}{[arXiv:2108.10323 [hep-th]]}.

\bibitem{CandelasHowManyCICYs}
P.~Candelas and A.~He, ``{On the number of complete intersection Calabi-Yau
  manifolds},'' Communications in Mathematical Physics {\bfseries 135} no.~1,
  (1990) 193 -- 199.

\bibitem{hubschCYB}
T.~H{\"u}bsch, {\em Calabi-Yau Manifolds: A Bestiary for Physicists}.
\newblock World Scientific, 1994.
\newblock \url{https://books.google.co.uk/books?id=Z5zSbFktn1EC}.

\bibitem{Grimm:2019bey}
T.~W. Grimm, F.~Ruehle, and D.~van~de Heisteeg, ``{Classifying
  Calabi\textendash{}Yau Threefolds Using Infinite Distance Limits},''
  \href{http://dx.doi.org/10.1007/s00220-021-03972-9}{Commun. Math. Phys.
  {\bfseries 382} no.~1, (2021) 239--275},
  \href{http://arxiv.org/abs/1910.02963}{[arXiv:1910.02963 [hep-th]]}.

\bibitem{TaylorJejjala}
V.~Jejjala, W.~Taylor, and A.~Turner, ``{Identifying equivalent Calabi--Yau
  topologies: A discrete challenge from math and physics for machine
  learning},'' in {\em {Nankai Symposium on Mathematical Dialogues}: {In
  celebration of S.S.Chern's 110th anniversary}}.
\newblock 2, 2022.
\newblock \href{http://arxiv.org/abs/2202.07590}{[arXiv:2202.07590 [hep-th]]}.

\bibitem{ConwaySloane}
J.~H. Conway and N.~J.~A. Sloane, ``Sphere packings, lattices and groups,''.
  \url{https://link.springer.com/book/10.1007/978-1-4757-6568-7}.

\bibitem{GaussBook}
C.~F. Gauss and A.~A. Clarke, {\em Disquisitiones Arithmeticae}.
\newblock Yale University Press, 1965.
\newblock \url{http://www.jstor.org/stable/j.ctt1cc2mnd}.

\bibitem{eichler1952quadratische}
M.~Eichler, {\em Quadratische Formen und Orthogonal Gruppen}.
\newblock Springer-Verlag, 1952.

\bibitem{LiE1992}
M.~A.~A. van Leeuwen, A.~M. Cohen, and B.~Lisser, {\em \texttt{LiE}, A package
  for Lie group computations}.
\newblock Computer Algebra Nederland, Amsterdam, 1992.

\bibitem{Cayley}
A.~Cayley, ``On the theory of linear transformations.'' Cambridge Mathematical
  Journal {\bfseries IV} (1845) 193--209.
  \url{www.archive.org/details/collectedmathema01cayluoft}.

\bibitem{GelfandDiscrim}
I.~Gelfand, M.~Kapranov, and A.~Zelevinsky, {\em Discriminants, Resultants, and
  Multidimensional Determinants}.
\newblock Modern Birkh{\"a}user Classics. Birkh{\"a}user Boston, 2008.
\newblock \url{https://books.google.co.uk/books?id=n6ZxAKevSbsC}.

\bibitem{AlgorithmForHdet}
A.~I. Barvinok, ``New algorithms for lineark-matroid intersection and
  matroidk-parity problems,'' Mathematical Programming {\bfseries 69} (1995)
  449--470. \url{https://api.semanticscholar.org/CorpusID:9422656}.

\bibitem{procesi2007lie}
C.~Procesi, {\em Lie Groups: An Approach through Invariants and
  Representations}.
\newblock Universitext. Springer New York, 2007.
\newblock \url{https://books.google.co.uk/books?id=Sl8OAGYRz_AC}.

\bibitem{DistribGCD}
P.~Diaconis and P.~Erdős, ``On the distribution of the greatest common
  divisor,'' Lecture Notes-Monograph Series {\bfseries 45} (2004) 56--61.
  \url{http://www.jstor.org/stable/4356298}.

\bibitem{Halverson:2019tkf}
J.~Halverson, B.~Nelson, and F.~Ruehle, ``{Branes with Brains: Exploring String
  Vacua with Deep Reinforcement Learning},''
  \href{http://dx.doi.org/10.1007/JHEP06(2019)003}{JHEP {\bfseries 06} (2019)
  003}, \href{http://arxiv.org/abs/1903.11616}{[arXiv:1903.11616 [hep-th]]}.

\bibitem{Cole:2019enn}
A.~Cole, A.~Schachner, and G.~Shiu, ``{Searching the Landscape of Flux Vacua
  with Genetic Algorithms},''
  \href{http://dx.doi.org/10.1007/JHEP11(2019)045}{JHEP {\bfseries 11} (2019)
  045}, \href{http://arxiv.org/abs/1907.10072}{[arXiv:1907.10072 [hep-th]]}.

\bibitem{Cole:2021nnt}
A.~Cole, S.~Krippendorf, A.~Schachner, and G.~Shiu, ``{Probing the Structure of
  String Theory Vacua with Genetic Algorithms and Reinforcement Learning},'' in
  {\em {35th Conference on Neural Information Processing Systems}}.
\newblock 11, 2021.
\newblock \href{http://arxiv.org/abs/2111.11466}{[arXiv:2111.11466 [hep-th]]}.

\bibitem{Constantin:2021for}
A.~Constantin, T.~R. Harvey, and A.~Lukas, ``{Heterotic String Model Building
  with Monad Bundles and Reinforcement Learning},''
  \href{http://dx.doi.org/10.1002/prop.202100186}{Fortsch. Phys. {\bfseries 70}
  no.~2-3, (2022) 2100186},
  \href{http://arxiv.org/abs/2108.07316}{[arXiv:2108.07316 [hep-th]]}.

\bibitem{Krippendorf:2021uxu}
S.~Krippendorf, R.~Kroepsch, and M.~Syvaeri, ``{Revealing systematics in
  phenomenologically viable flux vacua with reinforcement learning},''
  \href{http://arxiv.org/abs/2107.04039}{[arXiv:2107.04039 [hep-th]]}.

\bibitem{Abel:2021rrj}
S.~Abel, A.~Constantin, T.~R. Harvey, and A.~Lukas, ``{Evolving Heterotic Gauge
  Backgrounds: Genetic Algorithms versus Reinforcement Learning},''
  \href{http://dx.doi.org/10.1002/prop.202200034}{Fortsch. Phys. {\bfseries 70}
  no.~5, (2022) 2200034},
  \href{http://arxiv.org/abs/2110.14029}{[arXiv:2110.14029 [hep-th]]}.

\bibitem{Abel:2021ddu}
S.~Abel, A.~Constantin, T.~R. Harvey, and A.~Lukas, ``{String Model Building,
  Reinforcement Learning and Genetic Algorithms},'' in {\em {Nankai Symposium
  on Mathematical Dialogues}: {In celebration of S.S.Chern's 110th
  anniversary}}.
\newblock 11, 2021.
\newblock \href{http://arxiv.org/abs/2111.07333}{[arXiv:2111.07333 [hep-th]]}.

\bibitem{Abel:2022wnt}
S.~A. Abel and L.~A. Nutricati, ``{Ising Machines for Diophantine Problems in
  Physics},'' \href{http://dx.doi.org/10.1002/prop.202200114}{Fortsch. Phys.
  {\bfseries 70} no.~11, (2022) 2200114},
  \href{http://arxiv.org/abs/2206.09956}{[arXiv:2206.09956 [hep-th]]}.

\bibitem{Abel:2023zwg}
S.~Abel, A.~Constantin, T.~R. Harvey, A.~Lukas, and L.~A. Nutricati,
  ``{Decoding Nature with Nature's Tools: Heterotic Line Bundle Models of
  Particle Physics with Genetic Algorithms and Quantum Annealing},''
  \href{http://arxiv.org/abs/2306.03147}{[arXiv:2306.03147 [hep-th]]}.

\bibitem{Berglund:2023ztk}
P.~Berglund, Y.-H. He, E.~Heyes, E.~Hirst, V.~Jejjala, and A.~Lukas, ``{New
  Calabi-Yau Manifolds from Genetic Algorithms},''
  \href{http://arxiv.org/abs/2306.06159}{[arXiv:2306.06159 [hep-th]]}.

\bibitem{MyronenkoSong}
A.~Myronenko and X.~Song, ``Point set registration: Coherent point drift,''
  \href{http://dx.doi.org/10.1109/tpami.2010.46}{{IEEE} Transactions on Pattern
  Analysis and Machine Intelligence {\bfseries 32} no.~12, (Dec, 2010)
  2262--2275}. \url{https://doi.org/10.1109%2Ftpami.2010.46}.

\bibitem{pycpd}
S.~Khallaghi, ``pycpd: A pure numpy implementation of the coherent point drift
  cpd algorithm.'' \url{https://github.com/siavashk/pycpd}, 2016.

\bibitem{cytools}
M.~Demirtas, A.~Rios-Tascon, and L.~McAllister, ``{CYTools: A Software Package
  for Analyzing Calabi-Yau Manifolds},''
  \href{http://arxiv.org/abs/2211.03823}{[arXiv:2211.03823 [hep-th]]}.

\bibitem{Banchi_2015}
M.~Banchi, ``Rank and border rank of real ternary cubics,''
  \href{http://dx.doi.org/10.1007/s40574-015-0027-z}{Bollettino
  dell{\textquotesingle}Unione Matematica Italiana {\bfseries 8} no.~1, (Jul,
  2015) 65--80}. \url{https://doi.org/10.1007%2Fs40574-015-0027-z}.

\bibitem{CornellCountCY}
N.~Gendler, N.~MacFadden, L.~McAllister, J.~Moritz, R.~Nally, A.~Schachner, and
  M.~Stillman, ``{Counting Calabi-Yau Threefolds},''
  \href{http://arxiv.org/abs/2310.06820}{[arXiv:2310.06820 [hep-th]]}.

\bibitem{niven1991}
I.~Niven, H.~S. Zuckerman, and H.~L. Montgomery, {\em An Introduction to the
  Theory of Numbers}.
\newblock Wiley, New York, 5~ed., 1991.
\newblock Problem 16 of Section 3.5.

\bibitem{4FoldsGCDsJamesGray}
J.~Gray, A.~S. Haupt, and A.~Lukas, ``{Topological Invariants and Fibration
  Structure of Complete Intersection Calabi-Yau Four-Folds},''
  \href{http://dx.doi.org/10.1007/JHEP09(2014)093}{JHEP {\bfseries 09} (2014)
  093}, \href{http://arxiv.org/abs/1405.2073}{[arXiv:1405.2073 [hep-th]]}.

\bibitem{sl3cfundamentalinvariants}
M.~R. Bremner and J.~Hu, ``Fundamental invariants for the action of $sl_3(c)$
  × $sl_3(c)$ × $sl_3(c)$ on 3 × 3 × 3 arrays,'' Mathematics of Computation
  {\bfseries 82} no.~284, (2013) 2421--2438.
  \url{http://www.jstor.org/stable/42002748}.

\bibitem{deLaHarpeTopicsGeoGroup}
P.~de~la Harpe, {\em Topics in Geometric Group Theory}.
\newblock Chicago Lectures in Mathematics. University of Chicago Press, 2000.
\newblock \url{https://books.google.com.jm/books?id=cRT01C5ADroC}.

\end{thebibliography}\endgroup
\bibliographystyle{inspire}

\end{document}